\documentclass[twocolumn,showpacs,preprintnumbers,amsmath,amssymb,floatfix,prd]{revtex4}

\usepackage{graphicx}
\usepackage{amsmath}
\usepackage{amssymb}
\usepackage{amsfonts}
\usepackage{amscd}
\usepackage{mathrsfs}
\usepackage{psfrag}
\usepackage{subfigure}
\usepackage{float}
\usepackage{dcolumn}% Align table columns on decimal point
\usepackage{bm}% bold math

\newcommand{\eV}{\ensuremath{\mathrm{eV}}}
\newcommand{\diag}{\ensuremath{\mathrm{diag}}}

\newcommand{\I}{\ensuremath{\mathrm{i}}}
\DeclareMathOperator{\re}{Re}
\DeclareMathOperator{\im}{Im}

\def\beq{\begin{equation}}
\def\eeq{\end{equation}}
\def\be{\begin{equation}}
\def\ee{\end{equation}}
\def\bea{\begin{eqnarray}}
\def\eea{\end{eqnarray}}

\begin{document}
\received{09/18/2006}

\title{Quark lepton complementarity and renormalization group effects}
\author{Michael A. Schmidt}
\email{michael.schmidt@ph.tum.de}
\affiliation{Physik-Department T30, Technische Universit\"{a}t M\"{u}nchen, James-Franck-Stra{\ss}e, 85748 Garching, Germany}
\author{Alexei Yu. Smirnov}
\email{smirnov@ictp.trieste.it}
\affiliation{Physik-Department T30, Technische Universit\"{a}t M\"{u}nchen, James-Franck-Stra{\ss}e, 85748 Garching, Germany}
\affiliation{The Abdus Salam International Centre for Theoretical Physics, I-34100 Trieste, Italy}
\affiliation{Institute for Nuclear Research, Russian Academy of Science, Moscow, Russia}

\preprint{TUM-HEP-639/06}
\preprint{hep-ph/0607232}

\begin{abstract}
We consider a scenario for the Quark-Lepton Complementarity relations
between mixing angles in which the bi-maximal mixing
follows from the neutrino mass matrix.
According to this scenario in the lowest order the angle
$\theta_{12}$ is $\sim 1\sigma$ ($1.5 - 2^{\circ}$) above the best fit point
coinciding practically with the tri-bimaximal mixing prediction.
Realization of this scenario in the context of the seesaw type-I mechanism with leptonic
Dirac mass matrices  approximately equal to the quark mass  matrices is studied.
We calculate the renormalization group corrections to $\theta_{12}$
as well as to $\theta_{13}$ in the standard model (SM) and minimal supersymmetric standard model (MSSM).
We find that in large part of the parameter space corrections $\Delta \theta_{12}$
are small or negligible.
In the MSSM version of the scenario
the correction  $\Delta \theta_{12}$ is  in general
positive.  Small negative corrections  appear in
the case of an inverted mass hierarchy
and opposite CP parities of $\nu_1$ and $\nu_2$
when leading contributions to $\theta_{12}$ running are strongly suppressed.
The corrections are  negative in the SM version in a large part of the parameter
space  for values of the relative CP
phase of $\nu_1$ and $\nu_2$: $\varphi > \pi/2$.

\end{abstract}

\maketitle

\section{Introduction}
%%%%%%%%%%%%%%%%%%%%%%%%%%%%%%%%%%%%%%%%%%%%%%%%

Implications of the observed pattern of neutrino mass and mixing
(with two large angles) for fundamental physics
are still an open question. This pattern has
not yet led to a better  understanding of the origins of the neutrino mass
as well as fermion masses and mixing in general.
In contrast,  it made the situation more complicated
and more intriguing~\cite{mohsmi}.  In this connection, any hint from  data and
any empirical relation should be taken seriously and analyzed in details.

In fact,  one feature has been realized recently
that (if not accidental)  may lead to a
substantially  different  approach to the underlying physics.
Namely, the sums of the mixing  angles
of quarks and leptons for the 1-2 and 2-3 generations agree
with $45^{\circ}$ within $1\sigma$. In other words,
the quark and lepton mixings sum up to maximal mixing \cite{qlc,qlc1,qlc2}:
\be
\theta_{12} + \theta_C  \approx \frac{\pi}{4}, ~~~~~
\theta_{23} + \theta_{cb} \approx \frac{\pi}{4}.
%45^{\circ},
\label{qlcrel}
\ee
Here $\theta_C$ is the Cabibbo angle, $\theta_{cb} \equiv$ arcsin $V_{cb}$,
and $V_{cb}$ is the element of the Cabibbo-Kobayashi-Maskawa (CKM) matrix.
According to eqs. (\ref{qlcrel}) which are called
the quark-lepton complementarity (QLC) relations,
the quark and lepton mixings  are complementary to the maximal mixing.
(A  possibility that the lepton mixing
responsible for the solar neutrino conversion
equals maximal mixing minus $\theta_C$ was first
mentioned in \cite{petcov},  and
corrections to  the bimaximal mixing \cite{bm} from the CKM type rotations
were discussed in \cite{parametr}.)

For various reasons it is difficult to expect exact
equalities (\ref{qlcrel}). However certain correlations clearly show up:

\begin{itemize}

\item
the 2-3 leptonic mixing is close to maximal because
the 2-3 quark mixing, $V_{cb}$, is very small;

\item
the 1-2 leptonic mixing deviates from maximal
substantially because the 1-2 quark mixing ({\it i.e.}, the Cabibbo
angle) is relatively large.

\end{itemize}

If not accidental coincidence,  the QLC relations  imply~\cite{qlc1,qlc2,qlc-fm}

(i) a kind of quark-lepton symmetry
or quark-lepton unification which propagates the information about mixing from the
quark sector to the lepton sector.

(ii) existence of some additional structure which produces
maximal  or bi-maximal mixing.

Even within this context one expects some deviations
from exact quark-lepton complementarity due to

- broken quark-lepton symmetry,

- renormalization group (RG) effects.\\

There is a number of attempts to reproduce the QLC relations
on the basis of already  existing
ideas about fermion mass matrices
\cite{qlc1,qlc-fm,shift,qlc-km}.
Usually they lead to too small deviations of $\theta_{12}$
from $\pi/4$,  and therefore require further corrections
or deviations from the bi-maximal mixing or from the Cabibbo mixing
already in the lowest order.
So, in the majority of the models  proposed so far,  an approximate
QLC relation appears as a result of  an interplay of different independent
factors or as a sum of several independent contributions.
In these cases the QLC relation seems to be  accidental.
There are few attempts to construct a consistent
gauge model which explains the QLC relations.
The simplest possibility is the
$SU(2)_L \times SU(2)_R \times SU(4)_C$ model
that  implements the quark-lepton symmetry in the most straightforward
way \cite{qlc-fm,qlc-km}.
Phenomenology  of schemes with QLC relations has been extensively studied
\cite{qlc2,shift,qlc-cp,Hochmuth}.

The relation (\ref{qlcrel}) is realized at some high energy scale,
$M_F$,  of flavor physics and quark-lepton unification. Therefore
one should  take into account the  renormalization group effects
on the QLC relations when confronting them with the
low energy data. In fact,  it was marked in \cite{qlc2} that
in the Minimal Supersymmetric Standard Model (MSSM) the corrections are typically positive but
negative $\Delta \theta_{12}$ can be obtained from the RG effects
in presence of non-zero 1-3 mixing.
Also threshold corrections due to some
new scale of  physics, such as the low scale supersymmetry,  can
produce a negative shift of $\theta_{12}$, thus enhancing its  deviation from
$\pi/4$ \cite{qlc-ren}.

The Cabibbo mixing can be transmitted to the lepton sector in a more
complicated way (than via the quark-lepton symmetry). In fact, $\sin
\theta_C$ may turn out to be a generic parameter of the theory of
fermion masses - the ``quantum'' of flavor physics. Therefore it may appear
in various places: mass ratios, mixing angles.
One can consider the Cabibbo angle as an expansion parameter
for mixing matrices \cite{qlc2,parametr,parametr2,par-gen}.

In this paper we study in details the RG effects in the QLC scenario
where the bi-maximal mixing is generated by
the neutrino mass matrix. We calculate corrections
to the angles both in the Standard model (SM) and
MSSM. We analyze the dependence of the corrections
on various parameters and  obtain bounds on the parameters
from consistency condition with QLC. In particular, we
find regions where the corrections are negative.
The paper is organized as follows. In sec. 2 we
formulate the  scenario and  comment on parameterization dependence
of the QLC relations as well as  confront the relations with
experimental data. In sec. 3 we consider
realization of the scenario in the seesaw type I mechanism.
The RG effects in MSSM and SM are described  in secs. 4
and 5 correspondingly. We consider the RG effects on 1-3 mixing
and dependence of the effects
on scale of new physics in sec. 6. Conclusions are formulated in sec. 7.

\section{Update on  QLC}
%%%%%%%%%%%%%%%%%%%%%%%%%%%%%%%%%%%%%%%%%%%%%%%%%%%%%%%%%

\subsection{A scenario}
%%%%%%%%%%%%%%%%%%%%%%%%%%%%%%%%%%%%%%%%%%%%%%%%%%%%%%%%%%%%%%%%%%%%%%%%%%

A general scheme for the QLC relations is that
\be
``{\rm lepton~ mixing} =  {\rm bi\!-\!maximal~mixing} - {\rm CKM}'',
\ee
where the bi-maximal mixing matrix is~\cite{bm}
\be
U_{bm} = U_{23}^m U_{12}^m =
\frac{1}{2}
\left(\begin{array}{ccc}
\sqrt{2} & \sqrt{2} & 0\\
-1 & 1 & \sqrt{2}\\
1 & - 1 & \sqrt{2}
\end{array}
\right).
\label{bimax}
\ee
Here $U_{ij}^m$ is the  maximal mixing ($\pi/4$)  rotation in the $ij$-plane.

We assume that the bi-maximal mixing is generated by the neutrino
mass matrix.  That is, the same mechanism which is responsible for
the smallness of neutrino mass leads also to the large lepton mixing, and
it is the seesaw mechanism \cite{sees} that plays the role of additional structure
that generates the bi-maximal mixing. Therefore
\be
U_{PMNS} = U_l^{\dagger} U_{\nu} =  V_\mathrm{CKM}^{\dagger} \Gamma_{\alpha} U_{bm},
\label{qlc1mat}
\ee
where $\Gamma_{\alpha} \equiv \diag(e^{i\alpha_1}, e^{i\alpha_2}, e^{i\alpha_3})$
is the  phase matrix that can appear, in general,
at diagonalization of the charged lepton or neutrino Dirac mass matrices.

Similarity of the Dirac mass matrices
in the lepton and quark sectors, related to the quark-lepton symmetry,
is the origin of the CKM rotations in the lepton sector.
Here, there are two possibilities: \\

(i) In a certain (``symmetry'') basis, where the theory of flavor is formulated,
the neutrino  mass matrix is of the bi-maximal form. So  $U_{\nu} = U_{bm}$,
and the charged lepton mass matrix is diagonalized  by the CKM rotation:
\begin{equation}
U_l = V_\mathrm{CKM}.
\label{lckm}
\end{equation}
The problem here is that the masses
of charged leptons and down quarks are different: in particular,
$m_e/m_\mu = 0.0047$, whereas  $m_d/m_s = 0.04 - 0.06$,  and also
$m_\mu \neq  m_s$  at the grand unified theory (GUT) scale.
Since $m_l \neq m_d$, the equality (\ref{lckm}) implies
particular structure of the mass matrices  in which
mixing weakly depends on eigenvalues.

(ii) In the ``symmetry'' basis both the bimaximal and
CKM mixings come from the neutrino mass matrix, and
the charged lepton mass matrix is diagonal. That is, the
symmetry basis coincides with the flavor basis.
In this case the Dirac mass matrix of neutrinos is the origin
of the CKM rotation, whereas the Majorana mass matrix of
the right--handed (RH) neutrinos is responsible for the bi-maximal mixing.
Since the eigenvalues of the Dirac neutrino mass matrix are unknown
we can assume an exact equality of the mass matrices
\be
m_u = m^D_{\nu},
\label{uDmatr}
\ee
as a consequence of the quark-lepton symmetry.
%If the up-quark mass matrix, $m_u$,
%is the origin of the CKM mixing (and down quark matrix is diagonal),
The equality (\ref{uDmatr})
propagates the CKM mixing from the quark to the lepton sector
precisely. In this case, however, the Gatto-Sartori-Tonin relation between the
Cabibbo angle and  the  ratio of down quark  masses \cite{GST} turns out to be
accidental.
Furthermore,  one needs to explain why in the symmetry basis  both the charged
lepton and down quark mass matrices
are diagonal simultaneously in spite of  difference of eigenvalues.

These two cases have different theoretical implications,
however the phenomenological consequences and the RG effects are
the same. \\

In the scenario under consideration, the  relation (\ref{qlcrel}) is not realized precisely
even for zero phases $\alpha_i$ since the
$U_{12}^{CKM}$ rotation matrix should be permuted with $U_{23}^m$
in (\ref{qlc1mat}) to reduce the mixing matrix
to the standard parameterization form \cite{qlc2}.
From (\ref{qlc1mat}) we obtain the following expressions for the
leptonic mixing angles:
\begin{multline}
U_{e2} \equiv \cos\theta_{13} \sin \theta_{12} = \sin (\pi/4 -\theta_C)\\ +
0.5 \sin \theta_C \left[\sqrt{2} - 1 - V_{cb}
\cos (\alpha_3 - \alpha_1) \right]
%\nonumber
\\
 + 0.5 V_{ub}\cos (\alpha_3 - \alpha_1 -\delta_q),
\label{qlc12}
\end{multline}
where $\delta_q$ is the quark CP-violating phase.
This expression differs from the one derived in \cite{qlc2} by a factor
$\cos\theta_{13}$ as well as by the  last term, that  turn out to be relevant at the
level of accuracy we will consider here. The 1-3 mixing is large in this scenario
\cite{qlc2,qlc-cp,Hochmuth}:
\begin{multline}
\sin \theta_{13} = - \frac{\sin \theta_C}{\sqrt{2}} (1 - V_{cb} \cos \alpha_3) -
\frac{V_{ub}}{\sqrt{2}}\cos (\alpha_3  -\delta_q) \\\approx -
\frac{\sin \theta_C}{\sqrt{2}}
\label{qlc13}
\end{multline}
and, hence,  the Dirac CP phase $\delta$ is close to $180^\circ$.
So,  for the 1-2 mixing we find from (\ref{qlc12}) and  (\ref{qlc13})
\be
\sin \theta_{12} \approx U_{e2} (1 + \frac{1}{4}\sin^2 \theta_C),
\label{qlc12a}
\ee
and $U_{e2}$ is given in (\ref{qlc12}). Expression for the 2-3 mixing reads
\begin{multline}
U_{\mu 3} = \cos\theta_{13} \sin \theta_{12} \\=
\cos \theta_C \left[\sin (\pi/4 - \theta_{cb}) + \frac{V_{cb}}{\sqrt{2}}
(1 - \cos \alpha_3)
\right].
\label{qlc23}
\end{multline}
The RG effect on $V_\mathrm{CKM}$ is negligible.

%%%%%%%%%%%%%%%%%%%%%%%%%%%%%%%%%%%%%%%%%%%%%%%%%%%%%%%%%%%%%%
\subsection{QLC and parameterization.}
%%%%%%%%%%%%%%%%%%%%%%%%%%%%%%%%%%%%%%%%%%%%%%%%%%%%%%%%%%%%%%

The QLC relations are essentially  parameterization
independent. They can be expressed in terms of physical quantities (compare with
\cite{jarlskog}).
Indeed, the moduli of elements of the mixing matrix,
$U_{\alpha i}$,  are physical quantities immediately related to observables
and consequently,   parameterization independent.
In the standard parameterization
\be
|U_{e2}| = |\cos \theta_{13} \sin \theta_{12}|, ~~~
|U_{e3}|  = |\sin \theta_{13}|,
\ee
and therefore
\be
|\sin \theta_{12}| = \frac{1}{\sqrt{1 - |U_{e3}|^2}} |U_{e2}|.
\label{12expr}
\ee

%%Neglecting 1-3 mixing in the quark sector we

%There is some freedom of introduction of complex phase factor in the
%equality (\ref{}) (which in fact can account for some mismatch
%we will discuss later).

Notice that the presence of 1-3 mixing produces
some ambiguity in formulation of the QLC relations.
One can write the relations  in terms of angles
%%of 1-2 rotations
in the standard parameterization or in terms of matrix elements:
\be
\arcsin(V_{us}) + \arcsin(U_{e2}) = \pi/4.
\ee
Both forms coincide in the limit $U_{e3} \rightarrow 0$.

\subsection{Experimental status}
%%%%%%%%%%%%%%%%%%%%%%%%%%%%%%%%%%%%%%%%%%%%%%%%%%%%%%%%%%%%

\begin{figure}%[H]
\centerline{
{\includegraphics[width=8cm]{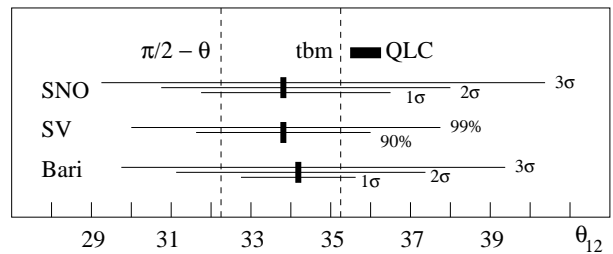}}\quad
}
\caption{The 1-2 mixing angle: experimental results and predictions.
Shown are the best fit values and allowed regions for $\theta_{12}$
from analyses of different groups SNO~\cite{sno}, Strumia--Vissani~\cite{sv}, Bari~\cite{bari}.
The vertical dashed lines correspond to value $(\pi/4 - \theta_C)$ and
tri-bimaximal mixing. The horizontal bar shows  values of QLC predictions
without RG corrections.
}
\label{fig1}
\end{figure}

In fig. \ref{fig1} we show results of determination of
$\theta_{12}$  by different
groups. SNO collaboration did analysis of the data in terms of
$2 \nu$ mixing \cite{sno}, whereas in \cite{sv} and \cite{bari} complete
$3\nu$ analyses have been performed.
Furthermore, in \cite{bari} a non-zero best fit  value of 1-3 mixing
has been obtained.
Results of different analyses are in a very good agreement:
\begin{align}
\theta_{12} = (33.8 \pm 2.2) ^{\circ} & (\theta_{13} = 0),\nonumber\\
\theta_{12} = (34.2 \pm 1.5)^{\circ} & (\theta_{13} = 7^{\circ}).
\label{bfval}
\end{align}
Notice that the determination of $\theta_{12}$ follows mainly from
 analysis of the solar neutrino data. In this analysis
$\theta_{12}$ and $\theta_{13}$ correlate.
In particular, the CC/NC ratio that gives the most
important restriction on mixing is determined by
$P \sim \cos^4 \theta_{13} \sin^2 \theta_{12}$. The best fit values
(\ref{bfval}) are along with the trajectory $P =$ constant.

%%%%%%%%%ffff0%%%%%%%%%%%%%%%%%%%%%%%%%%%%%%%%%%%%%%%%%%%%%%%%%%%%%
%%%%%%%%%%%%%%%%%%%%%%%%%%%%%%%%%%%%%%%%%%%%%%%%%%%%%%%%%%%%%%%%%%%%

In fig. \ref{fig1} we show the range  of QLC values of
$\theta_{12}$ obtained from eqs.  (\ref{qlc12},  \ref{qlc12a}) by varying $(\alpha_3
-
\alpha_1)$.  This variation gives
\be
\theta_{12}^{QLC} = 35.65 - 36.22^{\circ},
\label{qlcgrad}
\ee
or
\be
\sin^2 \theta_{12}^{QLC} = 0.340 - 0.349.
\label{qlcsin}
\ee
The smallest value of $\theta_{12}$ corresponds to
$\alpha_3 - \alpha_1 = -24^{\circ}$.

For the 1-3 mixing we obtain
\be
\sin^2 \theta_{13} \approx  \frac{1}{2} \sin^2 \theta_C  = 0.024,
\label{pred13}
\ee
which is at the upper  $1\sigma$ edge from the analysis \cite{bari}.

The sums of angles equal
\be
\theta_{12} + \theta_C = 46.7^{\circ} \pm 2.4^{\circ}  ~~~(1\sigma)
\ee
\be
\theta_{23} + \theta_{cb} = \left(43.9^{~+5.1}_{~-3.6} \right)^{\circ}
~~~(1\sigma).
\ee
The QLC prediction is slightly larger than the experimental
best fit point:
\be
\theta_{12}^{QLC} - \theta_{12}^{bf} = 1.5^{\circ} - 2.0^{\circ}.
\ee
The difference is well within $1\sigma$ of experimental measurements.
The exact complementarity value, $45^{\circ} - \theta_C$,
is $(1.8 -2.0)^{\circ}$  below the best fit value.
To disentangle these possibilities one needs to measure the 1-2 angle
with accuracy better than 1 degree:
$\Delta \theta_{12} < 1^{\circ}$,  that is translated into
\be
\frac{\Delta \sin^2 \theta_{12}}{\sin^2 \theta_{12}}
= \frac{2}{\tan\theta_{12}} \Delta \theta_{12} \sim
5\% (\Delta \theta_{12}/1^{\circ}),
\ee
or
\be
\frac{\Delta \sin^2 2\theta_{12}}{\sin^2 2\theta_{12}}
= \frac{4}{\tan 2\theta_{12}} \Delta \theta_{12} \sim
2.7\% (\Delta \theta_{12}/1^{\circ}).
\ee
Forthcoming results from SNO phase-III (He)  will improve
determination of the CC/NC ratio,  and consequently,
$\theta_{12}$.
Future low energy solar neutrino experiments aimed at measurements
of the pp-neutrino flux will have a   (1 - 2) \% sensitivity
to $\sin^2 2\theta_{12}$ provided that
degeneracy with 1-3 mixing is resolved.
Similar sensitivity could be achieved in  dedicated reactor
neutrino experiments with a large base-line \cite{12future}.

\subsection{QLC and tri-bimaximal mixing}
%%%%%%%%%%%%%%%%%%%%%%%%%%%%%%%%%%%%%%%%%%%%%%%%%%%%%%%%%%%%%%%%%

The QLC prediction (\ref{qlcsin})  is practically
indistinguishable from the tri-bimaximal
mixing \cite{tbm}  prediction  $\sin^2\theta_{12} = 1/3$.
So, it turns out  that almost the  same values of  1-2 mixing are obtained
from two different and independent combinations of matrices
\be
U_{23}^m U_{12}(\arcsin(1/\sqrt{3}))~~~~ {\rm and} ~~~~U_{12}(\theta_C)U_{23}^m
U_{12}^m.
\ee
There are two possible interpretations
of this fact:

The coincidence is  accidental,
which  means that one of the two approaches
(QLC or tri-bimaximal mixing) does not correspond to reality.
To some extend that can be tested by measuring the 1-3 mixing.
In the QLC-scenario one obtains (\ref{pred13}),
whereas the tri-bimaximal mixing implies  $\sin^2 \theta_{13} = 0$
unless some corrections are introduced.

The coincidence is not accidental, and therefore it implies a non-trivial
expression for the Cabibbo angle. Indeed,  from the equality $\sin \theta_{12}^{QLC} =
\sin \theta_{12}^{tbm}$ we obtain
\be
\sin \theta_C = \frac{2}{3\sqrt{3}} \left(\sqrt{\frac{5}{2}} -1 \right).
\label{cabibbo}
\ee

\section{Seesaw and QLC}
%%%%%%%%%%%%%%%%%%%%%%%%%%%%%%%%%%%%%%%%%%%%%%%%%%%%%%%%%%%%%%%%%%%

The seesaw mechanism \cite{sees} that provides a natural explanation of
smallness of neutrino mass
can also be  the origin of the  difference of the quark and lepton mixings,
and in particular, the origin of bimaximal
mixing. In the context of seesaw  type-I this implies a
particular structure of the RH neutrino mass matrix.

We assume that the type-I seesaw
gives the dominant contribution to neutrino masses
since it  can provide the closest relation
between the quark and lepton mass matrices, as required by the QLC.
The relevant terms of the Lagrangian are
\be
{\cal -L} = l^{cT} Y_e L H_d + N^T Y_{\nu} L H_u + \frac{1}{2}N^T M_R N
+ h.c.,
\label{lagran}
\ee
where $L \equiv (\nu, l)$ is the leptonic  doublet,  $N \equiv (\nu_R)^c$,
$H_d$ and $H_u$ are two different Higgs doublets in
the MSSM and $H_u = i\tau_2 H_d$ in SM;   $Y_e$ and $Y_{\nu}$ are the charged lepton and
neutrino Yukawa coupling matrices.
We consider $M_R$ as a bare mass matrix of the RH
neutrinos formed already  at the GUT or even higher scale. It can be generated by some
new interactions at  higher scales.

Decoupling of $N$ leads to the low energy effective $D=5$ operator
\be
\nu^T Y_{\nu}^T M_R^{-1} Y_{\nu} \nu H_u H_u.
\label{effop}
\ee
After the electroweak symmetry breaking this operator
generates the mass term for light neutrinos,
$\nu^T m_{\nu} \nu$  with
\be
m_{\nu} = - m_D^T M_R^{-1} m_D,
\ee
where  $m_D = Y_{\nu}\langle H_u \rangle$. \\

Let us consider the  basis where the neutrino Dirac mass matrix
is diagonal:
\be
Y_{\nu} = Y_{\nu}^{\diag} \equiv  \diag(y_1, y_2, y_3).
\ee
Then the light neutrino mass matrix equals
\be
m_{\nu} = - m_D^{\diag} M_R^{-1} m_D^{\diag},
\label{lightm0}
\ee
and $m_D^{\diag} \equiv Y_{\nu}^{\diag} \langle H_u \rangle$.

According to our assumption, the matrix (\ref{lightm0})
should generate the bimaximal rotation:
\be
m_{\nu} = m_{bm},
\ee
where in general,
\be
m_{bm} = \Gamma_{\delta} U_{bm} \Gamma_{\varphi/2}
m_{\nu}^{\diag} \Gamma_{\varphi/2}
U_{bm}^{T} \Gamma_{\delta}.
\label{bmmatr}
\ee
Here
\be
\Gamma_{\delta} \equiv \diag(e^{\I\delta_1}, e^{\I\delta_2}, e^{\I\delta_3}),
\ee
is the phase matrix,
\be
m_{\nu}^{\diag} \equiv \diag(m_1, m_2, m_3)
\ee
is the diagonal matrix of the light neutrinos, and
\be
\Gamma_{\varphi} \equiv \diag(e^{\I\varphi_1 /2}, e^{\I\varphi_2 /2}, 1),
\ee
with $\varphi_i$  being the Majorana phases of light neutrinos. \\

According to our assumption, the CKM rotation follows from diagonalization of the
charged lepton matrix
\be
V_\mathrm{CKM}^\dagger Y_e^\dagger Y_e V_\mathrm{CKM} =
\diag\left(y_e^2,\,y_\mu^2,\,y_\tau^2 \right),
\ee
and we will parameterize it as
\be
V_l =  \Gamma_\phi V_\mathrm{CKM} (\theta_q, \delta_q).
\label{CKMpar}
\ee
Here the diagonal matrix of the phase factors on the RH side
has been  absorbed in the charged lepton field redefinition;
$V_\mathrm{CKM}$ is  the CKM matrix in the standard parameterization,
$\theta_q$ and  $\delta_q$ are the quark (CKM) mixing angles, and
\be
\Gamma_\phi \equiv \diag(e^{\I\phi_1}, e^{\I\phi_2}, e^{\I\phi_3}).
\label{phasematr}
\ee
Thus, in general, there are three matrices of phases,
$\Gamma_{\delta}$,  $\Gamma_{\varphi}$ and  $\Gamma_{\phi}$, relevant for
relations between the mixing angles.
Finally, from
(\ref{CKMpar}) and (\ref{bmmatr}) we obtain
\be
U_{PMNS} = V_\mathrm{CKM}^{\dagger} (\theta_q, \delta_q)
\Gamma (\delta_l - \phi_l) U_{bm},
\ee
and therefore in (\ref{qlc1mat})
$\alpha_j = (\delta_j - \phi_j)$. \\

The neutrino mass matrix in the flavor basis equals
\be
m_{\nu}^f = V_\mathrm{CKM}^{T} m_{bm} V_\mathrm{CKM}.
\ee

From (\ref{lightm0}) and (\ref{bmmatr})
we find an expression for the RH neutrino mass matrix:
\be
M_{R}  =  \Gamma_\delta m_D^{\diag}
U_{bm}
\Gamma_{\varphi/2} (m_{\nu}^{\diag})^{-1} \Gamma_{\varphi/2}
U_{bm}^{T} m_D^{\diag}
\Gamma_{\delta}.
\label{RHmatrix}
\ee
Omitting the phase factor $\Gamma (\delta_i)$
(that can be absorbed in the definition of $M_{R}$) and including
the CP phases $\varphi_i$ into masses of light neutrinos
$\Gamma_{\varphi/2} (m_{\nu}^{\diag})^{-1} \Gamma_{\varphi/2} = (m_{\nu,
\varphi}^{\diag})^{-1}$, we obtain
\be
M_{R}  =  m_D^{\diag}U_{bm}
(m_{\nu, \varphi}^{\diag})^{-1} U_{bm}^{T} m_D^{\diag}.
\ee
Explicitly
\be
M_R = \frac{1}{4}m_D^{\diag}
\begin{pmatrix}
2A  & -\sqrt{2} B & -\sqrt{2} B\\
... & C + A & C - A\\
... & ... &   C + A\\
\label{rhmatr}
\end{pmatrix}
m_D^{\diag},
\ee
where
\be
A \equiv \frac{1}{m_1} + \frac{1}{m_2}, ~~~B \equiv \frac{1}{m_2} -  \frac{1}{m_1},
~~~
C \equiv \frac{2}{m_3},
\label{abc}
\ee
(with phases $\varphi_i$ included).
We can parameterize $m_D^{\diag}$ as
\be
m_D^{\diag} =  m_t \diag(\epsilon'^{2}, \epsilon, 1),
\ee
with  $m_t$ being the mass of top quark and
$\epsilon' \approx \epsilon \sim 3 \cdot 10^{-3}$.
Using  smallness of $\epsilon$'s it is easy to estimate the mass eigenvalues:
\begin{multline}
M_3 \approx \frac{m_t^2}{4} (A + C), ~~
M_2 \approx m_t^2 \epsilon^2 \frac{AC}{A + C},\\
M_1 \approx m_t^2 \epsilon'^4 \frac{A^2 - B^2}{2A}.
\label{123mb}
\end{multline}
Furthermore, the 1-2 and 2-3 mixing angles are of the order $\epsilon$,
whereas 1-3 mixing is of the order $\epsilon^2$.

In the case of normal mass hierarchy, $m_1 \ll m_2 \ll m_3$,
eqs. (\ref{123mb}) lead to
\be
M_3 \approx \frac{m_t^2}{4m_1}, ~~~
M_2 \approx \frac{2 m_t^2 \epsilon^2}{m_3}, ~~~
M_1 \approx \frac{2 m_t^2 \epsilon'^{4}}{m_2},
\label{123ma}
\ee
in agreement with results of \cite{AFS}.
Notice a permutation character of these expressions:
the masses of RH neutrinos 1, 2, 3 are determined by light masses 2, 3, 1.
With $m_1 \rightarrow 0$, apparently,
$M_3 \rightarrow \infty$.
For $\epsilon' =  \epsilon \sim 3 \cdot 10^{-3}$  and $m_1 = 10^{-3}$ eV
values of masses equal
\be
M_3  =  9 \cdot 10^{15} ~{\rm GeV},~~
M_2 =   1 \cdot 10^{10} ~{\rm GeV},~~
M_1 =  5 \cdot 10^{5} ~~{\rm GeV}.
\ee
So,  masses have a ``quadratic'' hierarchy.\\

In the case of inverted mass hierarchy, $m_3 \ll m_1 \approx m_2 \equiv m_A$,  and the
same CP  phases of $\nu_1$ and  $\nu_2$ we obtain from (\ref{123mb})
\be
M_3 \approx \frac{m_t^2}{4m_3}, ~~~
M_2 \approx \frac{2 m_t^2 \epsilon^2}{m_A}, ~~~
M_1 \approx \frac{2 m_t^2 \epsilon'^{4}}{m_A},
\label{123maI}
\ee
where $m_A \equiv \sqrt{|\Delta m^2_{31}|}$.
This  leads again to a strong mass hierarchy.
Notice that now the mass of the lightest RH neutrino
is determined by the atmospheric mass scale.
Thus, apart from special regions in the parameter
space that  correspond to level crossings (see sect. 5) the QLC
implies generically a very strong (``quadratic'') mass hierarchy of the RH
neutrinos and very small mixing: $\Theta_{ij} \sim \epsilon$.
As we will see, this determines substantially the size of the RG effects.

Let us introduce the unitary matrix, $U_R$, which diagonalizes the right--handed
neutrino mass matrix
\be
 U_R^T M_R U_R =  M_R^{\diag} \equiv  \diag\left(M_1,\,M_2,\,M_3\right),
\label{rphases1}
\ee
and  the mixing matrix can be parameterized as
\be
U_R = \Gamma_{\Delta} V_{CKM} (\Theta_{ij}, \Delta) \Gamma_{\xi/2},
\label{rphases2}
\ee
where $\Theta_{ij}$ and $\Delta$ are the angles  and CP-phase of the RH neutrino
mixing matrix.\\

In what follows we will not elaborate further on the origin of
particular structures of  $M_R$ (\ref{rhmatr}), just noticing that it  can be related
to the double (cascade)
seesaw mechanism \cite{dss} with the ``screening'' of Dirac structure \cite{scre,kim}.
%We will discuss the RG effects.

\section{RG effects: general consideration and  the MSSM case}
%%%%%%%%%%%%%%%%%%%%%%%%%%%%%%%%%%%%%%%%%%%%%%%%%%%%%%%%%%%%%%%%%

\subsection{General consideration}
%%%%%%%%%%%%%%%%%%%%%%%%%%%%%%%%%%%%%%%%%%%%%%%%%%

The quark-lepton symmetry  implied by the QLC relation
means that physics responsible for this
relations should be realized at some scale $M_F$ which is  at the
quark-lepton unification scale, $M_{GUT}$,
or  even  higher scales. An alternative possibility would be
the quark-lepton relation due to the  Pati-Salam symmetry \cite{pati}
broken below the GUT scale.
Consequently,   there are three different regions of RG running:

(i) below the seesaw scales, $\mu < M_1$,  where $M_1$ is the lightest RH neutrino mass.
In this region all three neutrinos decouple and the D=5 operator
(\ref{effop}) is formed;

(ii) between  the seesaw scales, $M_1  < \mu < M_3$,  where $M_3$
is the heaviest RH neutrino mass;

(iii) above the seesaw scales  $ M_3 <  \mu < M_F$.
If  $M_F > M_{GUT}$ new features of running can appear above $M_{GUT}$.\\

The RG equation for the neutrino mass matrix is given by~\cite{rgeMat,rge-eq,massmatrrg}
\be
16 \pi^2 \Dot m_{\nu} = P^T m_{\nu} + m_{\nu}P  + \kappa m_{\nu},
\label{eqrun}
\ee
where $\Dot m_{\nu} \equiv  \mu\, \mathrm{d} m_{\nu}/\mathrm{d} \mu$, $\mu$ is the renormalization scale,
and $\kappa m_{\nu}$ includes the  gauge interaction terms that can influence
the flavor structure in the SM case (see below);
\be
P \equiv C_e Y_e^\dagger Y_e + C_\nu Y_\nu^\dagger Y_\nu,
\label{defP}
\ee
$C_e = -3/2$, $C_\nu = 1/2$ in the SM and $C_e = C_\nu = 1/2$ in the MSSM.
From  the  evolution equation (\ref{eqrun})
for the mass matrix we can obtain
the equations~\footnote{The renormalization group equations
of the mixing angles are taken from~\cite{exactRGE, aboveseesaw}}  for
observables (masses and mixings) \cite{manfth,exactRGE,manf13,aboveseesaw}.
Above and below the seesaw scales the gauge interactions
produce a flavor universal effect and  all contributions (from
all RH neutrinos)
to the  neutrino mass matrix have the same renormalization group equation.

In the limit of vanishing 1-3 mixing the evolution of $\theta_{12}$ is described
approximately by
\begin{widetext}
\begin{multline}
32\pi^2
\Dot\theta_{12}=\mathcal{Q}_{12}\left[\sin 2\theta_{12}\left(P_{11}-
c_{23}^2P_{22}-s_{23}^2P_{33} + s_{23}\re
    P_{23}\right) +   \right.\\\left.  + 2\cos 2\theta_{12}\left(c_{23}\re
P_{21}-s_{23} \re
    P_{31}\right)\right] + 4\mathcal{S}_{12}\left(c_{23}\im P_{21}-s_{23}\im
  P_{31}\right) ,
\label{eq:DotTheta12}
\end{multline}
\end{widetext}
where  $s_{23} \equiv \sin \theta_{23}$, $c_{23} \equiv  \cos \theta_{23}$, {\it etc.},
\be
\mathcal{Q}_{ij}  \equiv  \frac{|m_i e^{\I\varphi_i} +
m_j e^{\I\varphi_j}|^2}{\Delta m_\mathrm{ji}^2},
\ee
%%where $\tilde m_i=m_i e^{\I\varphi_i}$ are the masses including Majorana phases
and
\be
\mathcal{S}_{12} \equiv \frac{m_1 m_2
  \sin\left(\varphi_1-\varphi_2\right)}{\Delta m^2_\mathrm{21}}.
\ee
Above the seesaw  scale one needs to consider renormalization of
couplings  of the full Lagrangian (\ref{lagran}).
The evolution  of the effective operator which gives  masses to neutrinos
after the electroweak symmetry breaking
is  determined  by evolution of  the neutrino Yukawa couplings $Y_{\nu}$
and the mass terms of right--handed neutrinos.

Below the seesaw scales, running is dominated by $P_{33}$ in the flavor basis
which results  in an increase of $\theta_{12}$
in  MSSM and a slight decrease in the SM
due to different signs of $C_e$:
\begin{equation}
32\pi^2 \Dot\theta_{12} \approx -\mathcal{Q}_{12}\sin 2\theta_{12} s_{23}^2P_{33}\; .
\end{equation}

Above the seesaw scales, the leading contribution is again given by $P_{33}$,
and the next--to--leading contribution comes from
$P_{32}$. This yields an increase of $\theta_{12}$
when running to  low scales both in the  MSSM and in
SM. Explicitly the corresponding evolution equation can be written as
%%%
\begin{multline}
32\pi^2\Dot\theta_{12}=-\mathcal{Q}_{12}  C_\nu \sin 2\theta_{12}
\sin \theta_{23}\\\left[\sin \theta_{23} -  V_{cb}
  \cos \theta_{23} \cos\left(\phi_2 - \phi_3\right)\right]\; ,
\label{evol12a}
\end{multline}
%%%%
%%where $A_q$, $\lambda = \sin \theta_C$,  are the Wolfenstein parameters of the CKM
%%matrix $V_\mathrm{CKM}$ and $\phi_\mu$ and $\phi_\tau$
%%are the phases defined by
%%$V=\diag(e^{i \phi_e},\,e^{i \phi_\mu},\,e^{i \phi_\tau}) V_\mathrm{CKM}$. \\
and the phases $\phi_i$ are determined in Eq. (\ref{CKMpar}, \ref{phasematr}).

Effect of running between the seesaw scales (about 10-orders of magnitude in
$\mu$)  is  more complicated. In this range  the Yukawa coupling matrix has two
terms (contributions):

(1) D=5 effective operators for the light neutrinos formed after decoupling of one or
two RH neutrinos,

(2) Dirac type couplings  and mass terms for undecoupled RH neutrinos
given by the Lagrangian (\ref{lagran}).

These terms  are renormalized differently. In particular, for the terms of second type
the neutrino Yukawa couplings are important.
The difference, however, cancels (between the seesaw scales) in the case of MSSM
in which  only the wave function renormalization
takes place due to the non--renormalization theorem. In contrast, in the  SM due to
vertex corrections to the D=5 operators
the difference does not cancel, and, as we will see,
produces a significant effect.
So, in the MSSM, the RG equations are the same for both contributions and
eq.~\eqref{eq:DotTheta12} is valid. In the SM they are not equal and
eq.~\eqref{eq:DotTheta12} can not be applied.\\

After the heaviest right--handed neutrino is
integrated out, the right--handed neutrino mixing at the threshold influences
running of $\theta_{12}$. In the second order of
$\sin \theta_C$, the expression for $\Dot\theta_{12}$ reads:
\begin{multline}
32\pi^2\Dot\theta_{12}=-\frac{1}{4}\mathcal{Q}_{12} C_\nu \left(s_{23} -
  V_{cb} c_{23} \cos\left(\phi_2 -\phi_3\right)\right)\\
\left(3 - 2\cos 2\Theta_{23}\cos^2\Theta_{13}-
\cos 2\Theta_{13}\right)\sin2\theta_{12}
s_{23} \; ,
\label{evol12b}
\end{multline}
where $\Theta_{ij}$ are the right-handed neutrino mixing angles
at the scale at which  the heaviest neutrino is integrated out.
The unitary rotation of the right--handed neutrino fields
is done at the threshold of the heaviest
right--handed neutrino,   and the exact definitions of
the angles are given in Eq. (\ref{rphases1}, \ref{rphases2}).

\subsection{RG evolution and scales of flavor physics}
%%%%%%%%%%%%%%%%%%%%%%%%%%%%%%%%%%%%%%%%%%%%%%%%%%%%%%%%%%%%%%%%%%%%%

We have performed running from the $M_F$ scale
down to the electroweak  scale  and calculated
$\Delta\theta_{12} \equiv \theta_{12}(M_Z)-\theta_{12}(M_\mathrm{F})$.
For that we solved numerically a complete set of the RG equations including sub-leading
effects due to the non-zero 1-3 mixing.
In most of our calculations we take for definiteness
$M_F = M_{GUT} = 2 \cdot 10^{16}$ GeV. We consider  separately
dependence of the results on $M_F$.
Notice that the renormalization of $\theta_{12}$ in the bimaximal
scheme has been studied in
\cite{bmrad}.

The following free parameters  determine the RG effects substantially:
the absolute scale of light masses $m_1$, the CP (Majorana) phases
of light neutrinos, $\varphi_i$,  and the phases $\alpha_i$.
We studied dependence of the RG effects on these parameter.
For each set of the parameters we have calculated  the RH
mass matrix and  running effects.
The angles are fixed by the QLC relation at $M_{F}$,
and the mass squared differences
are  adjusted to lie in the experimentally allowed region
at the electroweak scale.
For the neutrino  Yukawa couplings we take
$y_1:y_2:y_3= \epsilon^2: \epsilon:1$,
($\epsilon = \epsilon'$)  and  $\epsilon=3 \cdot 10^{-3}$.

We consider  RG evolution  in the MSSM with a unique
SUSY threshold 1 TeV.
The RG effects depend on the absolute mass scale, $m_1$,
$\tan \beta$, and the relative phase between the
first and second mass eigenstates $\varphi \equiv \varphi_2 - \varphi_1$.
Dependence on  other
parameters (e.g., other phases) is rather weak.
Still we will use explicitly the phase $\varphi_2$ keeping everywhere $\varphi_1 = 0$.

In what follows  we will describe results of our
numerical calculations. We  give an interpretation
of the results using approximate
formulas presented in Secs. III and IV.1.

\subsection{RG effect in MSSM with normal mass hierarchy}
%%%%%%%%%%%%%%%%%%%%%%%%%%%%%%%%%%%%%%%%%%%%%%%%%%

%We study first the case normal mass hierarchy
%of light neutrinos.

In fig. \ref{fig:MSSMTheta12}  we show some examples of the scale dependence of
$\theta_{12}$ for various values of parameters.
With increase of $m_1$   two factors enhance
the RG effects:

%%%%%%%%ffff1ab%%%%%%%%%%%%%%%%%%%%%%%%%%%%%%%%%%%%%%%%%%%%%%%%%%%%%%%%%%%%%%%%%%
\begin{figure*}
\centerline{
\subfigure[~]
{\label{fig:MSSMTheta12a}\includegraphics[width=7.5cm]{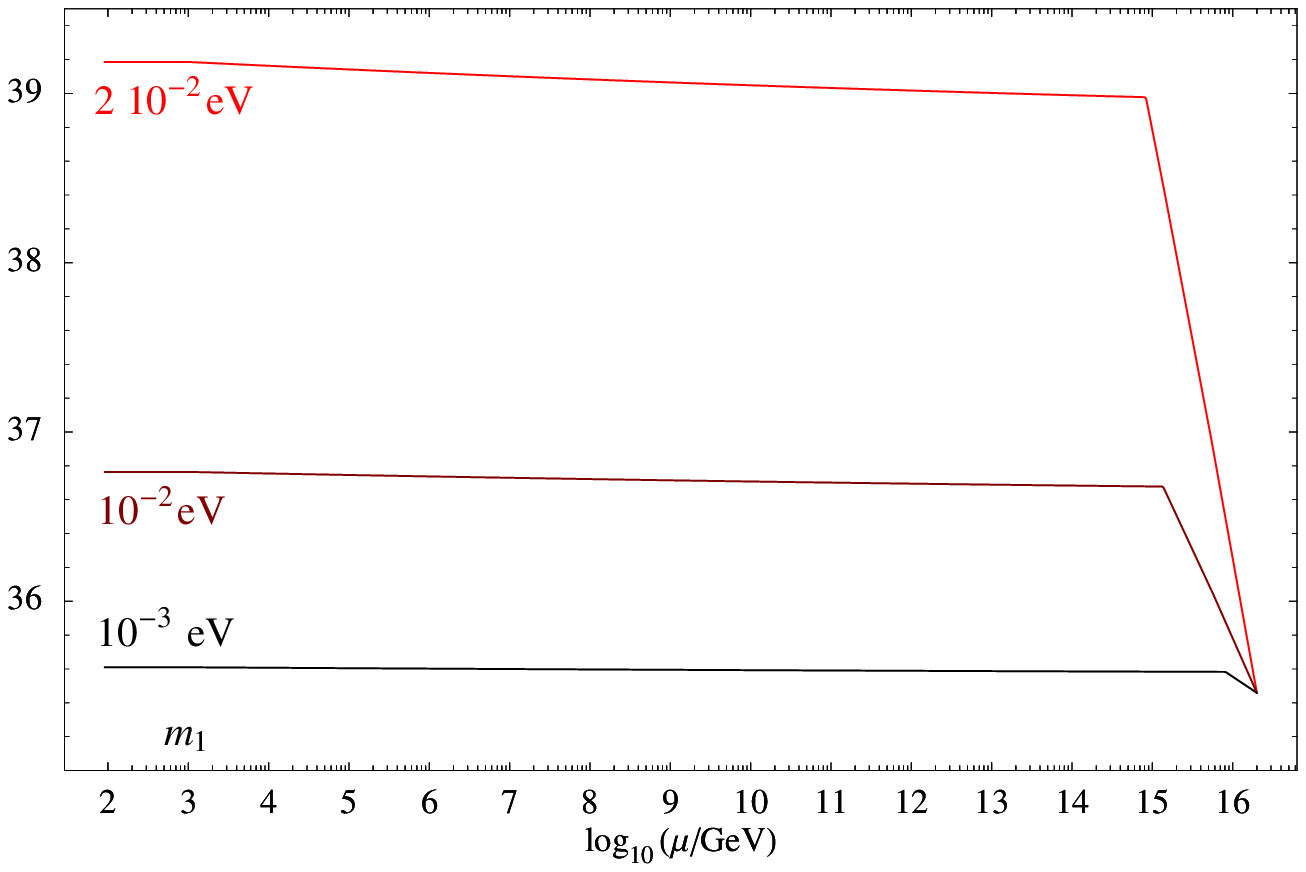}}\quad
\subfigure[~]
{\label{fig:MSSMTheta12b}\includegraphics[width=7.5cm]{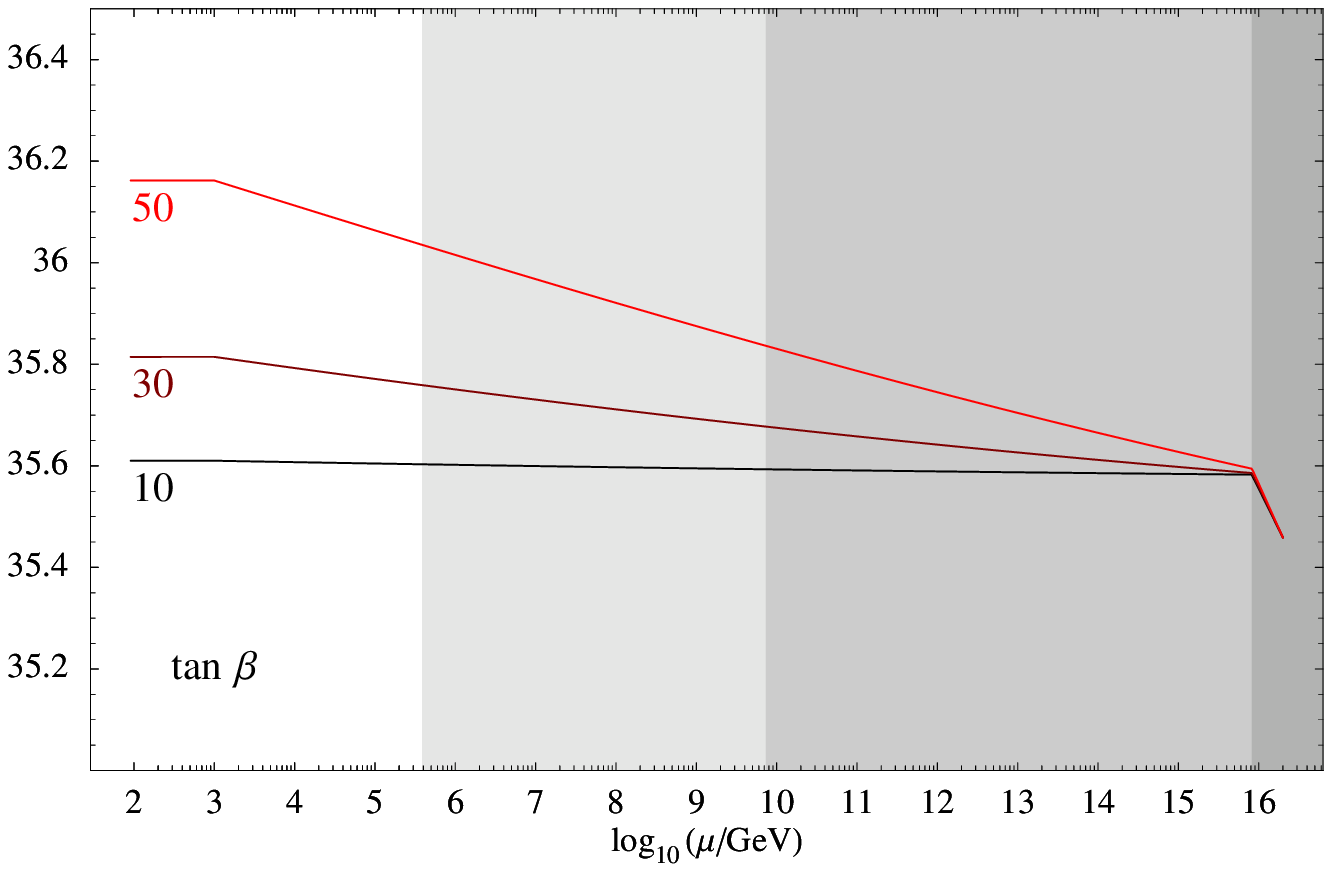}}}
\caption{Examples of running of $\theta_{12}$ in the case of MSSM and normal mass
hierarchy. The dependence of $\theta_{12}$ on $\mu$ (a) for different
values of $m_1$, and   $\tan \beta = 10$;
(b) on $\tan \beta$ for $m_1 = 10^{-3}$ eV.
All the CP-phases are taken to be zero.}
\label{fig:MSSMTheta12}
\end{figure*}
%%%%%%%%%%%%%%%%%%%%%%%%%%%%%%%%%%%%%%%%%%%%%%%%%%%%%%%%%%%%%%%%%%%%%%%%%%%%%%%%%%

(1) the  largest mass $M_3$ decreases according to (\ref{123ma}).
Correspondingly, the region  above
the seesaw scale, $M_3 - M_{GUT}$ increases;

(2) corrections to the mass matrix elements are proportional to  values of
elements:
$\Delta m_{\alpha\beta} \propto m_{\alpha\beta}$ and  since with increase
of $m_1$  the masses,  $m_{\alpha\beta}$, generically increase, the corrections
increase correspondingly.

For relatively small  $\tan \beta \sim (3 - 10)$,  the
dominant  contribution follows from region above the seesaw
scales due to large $(Y_{\nu})_{33}$. Evolution below $M_3$ is
mainly due to the Yukawa couplings $Y_e$ which
are relatively small. The effect increases fast with $m_1$:
\be
\Delta \theta_{12} \propto Q_{12} \log (M_{GUT}/M_3).
\ee
Notice that
$M_3 \propto 1/m_1$. Therefore for
$m_1 \sim 10^{-3}$ eV  the running of $\theta_{12}$ is mainly related to
increase of region above the seesaw scale.
For $m_1 > 10^{-2}$ eV the spectrum of light neutrinos
becomes degenerate and $\Delta \theta_{12} \propto Q_{12}
\propto m_1^2$ (fig. \ref{fig:MSSMTheta12a}).
For large $\tan \beta$ and small $m_1$ the dominant
contribution to $\Delta \theta_{12}$ comes from the region below $M_3$
where $\Delta \theta_{12} \propto \tan^2 \beta$
(see fig. \ref{fig:MSSMTheta12b}).

%%%%%%%%%%%%%%%%%%%%%%%%%%%%%%%%%%%%%%%%%%%%%%%%%%%%%%%%%%%%%%
\begin{figure}
\centerline{
{\includegraphics[width=6.5cm]{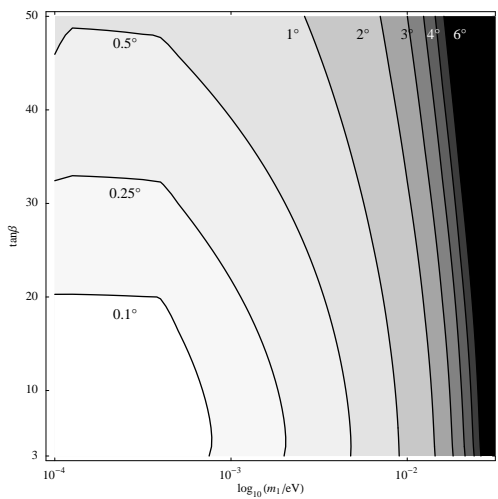}}\quad
}
\caption{Contours of constant RG corrections, $\Delta \theta_{12}$, in the
$\tan \beta - m_1$ plane in the case  of  MSSM and normal mass hierarchy.
All the CP-phases are taken to be zero.}
\label{fig:MSSMtanBetaM1}
\end{figure}
%%%%%%%%%%%%%%%%%%%%%%%%%%%%%%%%%%%%%%%%%%%%%%%%%%%%%%%%%%%%%%%%%%%%

A combined  dependence of corrections, $\Delta \theta_{12}$, on
$m_1$ and  $\tan\beta$ is presented in fig. \ref{fig:MSSMtanBetaM1} where we
show contours of constant $\Delta \theta_{12}$ in
the $(m_1 - \tan\beta)$ plane. The change of behavior of
contours at $m_1 = 8 \cdot 10^{-4}$ eV is a
consequence of our boundary condition: At $m_1 < 8 \cdot 10^{-4}$ eV
we have  $M_3 > M_{GUT}$,  and therefore the region above seesaw scale disappears.

%%%%%%%%%%%%%%%%%%%%%%%%%%%%ffff4%%%%%%%%%%%%%%%%%%%%%%%%%%%%%%%%%%%%%%%%%
\begin{figure}
  \centerline{
{\includegraphics[width=8cm]{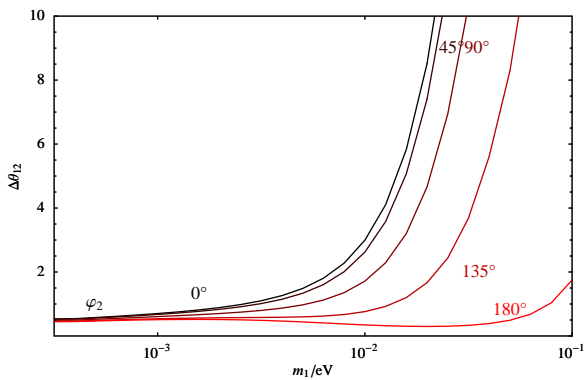}}\quad
}
\caption{The dependence of the RG correction, $\Delta \theta_{12}$ (in degrees),
on $m_1$  for different values of $\varphi_2$ (figures at the curves) in the MSSM  with the
normal mass  hierarchy. The lines correspond to
$\tan\beta=10$ and $\varphi_1 = 0$.
}
\label{fig:MSSMPhi2L}
\end{figure}
%%%%%%%%%%%%%%%%%%%%%%%%%%%%%%%%%%%%%%%%%%%%%%%%%%%%%%%%%%%%%%%%%%%%%%%%%%%%%%%%%

In fig. \ref{fig:MSSMPhi2L} we show the correction
$\Delta \theta_{12}$ as functions of $m_1$ for different values of
$\varphi_2$. The dependence of $\Delta \theta_{12}$  on $\varphi_2$,
given essentially   by the factor $Q_{12}$,
is weak for the hierarchical spectrum,
$m_1 \ll 8\cdot 10^{-3}$ eV, and very strong for
the degenerate spectrum: $\Delta \theta_{12} \propto
(1 + \cos \varphi)$. The corrections are strongly suppressed
for the opposite CP-parities, $\varphi_2 = 180^{\circ}$,
(fig. \ref{fig:MSSMPhi2L}) that  agrees with the results of
previous studies of corrections in  the
degenerate case \cite{degen,rad1,renphases,small}.\\

Corrections $\Delta \theta_{12}$ are positive.
This fact is essentially
a consequence of strong hierarchy of the Yukawa
couplings $Y_{\nu}$ and $Y_e$. The evolution is
given approximately by eq.~(\ref{eq:DotTheta12}), where
$P_{33} \propto 1/2 (|(Y_e)_{33}|^2 +  |(Y_{\nu})_{33}|^2) > 0$.
The off-diagonal couplings $P_{ij}$ are much smaller.
Since $Q_{12} > 0$ we obtain  $\dot{\theta}_{12} < 0$,   that is, the angle
$\theta_{12}$ increases with decrease of $\mu$.

Condition  that the QLC prediction for $\theta_{12}$
is within $1\sigma$ of the best fit experimental value requires
$\Delta \theta_{12} < 0.5^{\circ} - 1^{\circ}$. This, in turn,  leads to
bounds on parameters of neutrino spectrum and $\tan\beta$.
In particular, according to fig. \ref{fig:MSSMPhi2L} the degenerate neutrino
spectrum is excluded for the same CP parities ($\varphi_2 = 0$).
In the case of large $\tan\beta$ it requires strongly hierarchical
spectrum: $m_1 < 10^{-3}$ eV that eliminates the running region above seesaw scale.
However, a degenerate spectrum  is allowed for $\varphi \sim 180^{\circ}$.

Taking  $2\sigma$ upper bound  $\Delta \theta_{12} < 2^{\circ}$
we find that the quasi-degenerate spectrum with $m_1 \sim 10^{-2}$ eV
is allowed even for the same parities.
For normal mass hierarchy with $m_1 <  10^{-3}~ \eV$ and
$\tan\beta \sim (3 - 10)$ the running effect is negligible:
$\Delta \theta_{12} < 0.1^{\circ}$.\\

%%%%%%%%ffff5ab%%%%%%%%%%%%%%%%%%%%%%%%%%%%%%%%%%%%%%%%%%%%%%%%%%%%%%%%%%%%%%%%%%
\begin{figure*}
\centerline{
\subfigure[~]
{\label{fig:MSSMTheta12Ia}\includegraphics[width=7.5cm]{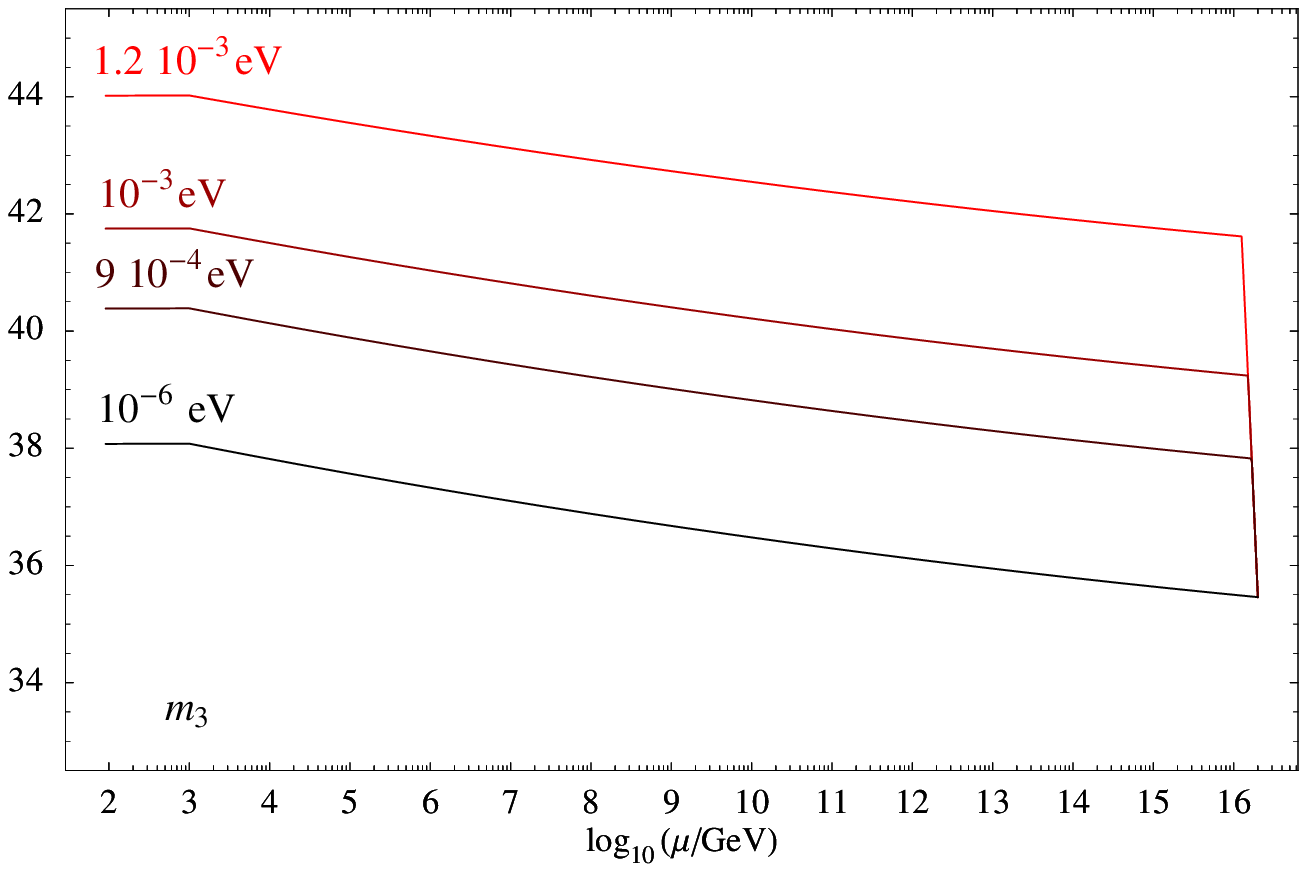}}\quad
\subfigure[~]
{\label{fig:MSSMTheta12Ib}\includegraphics[width=7.5cm]{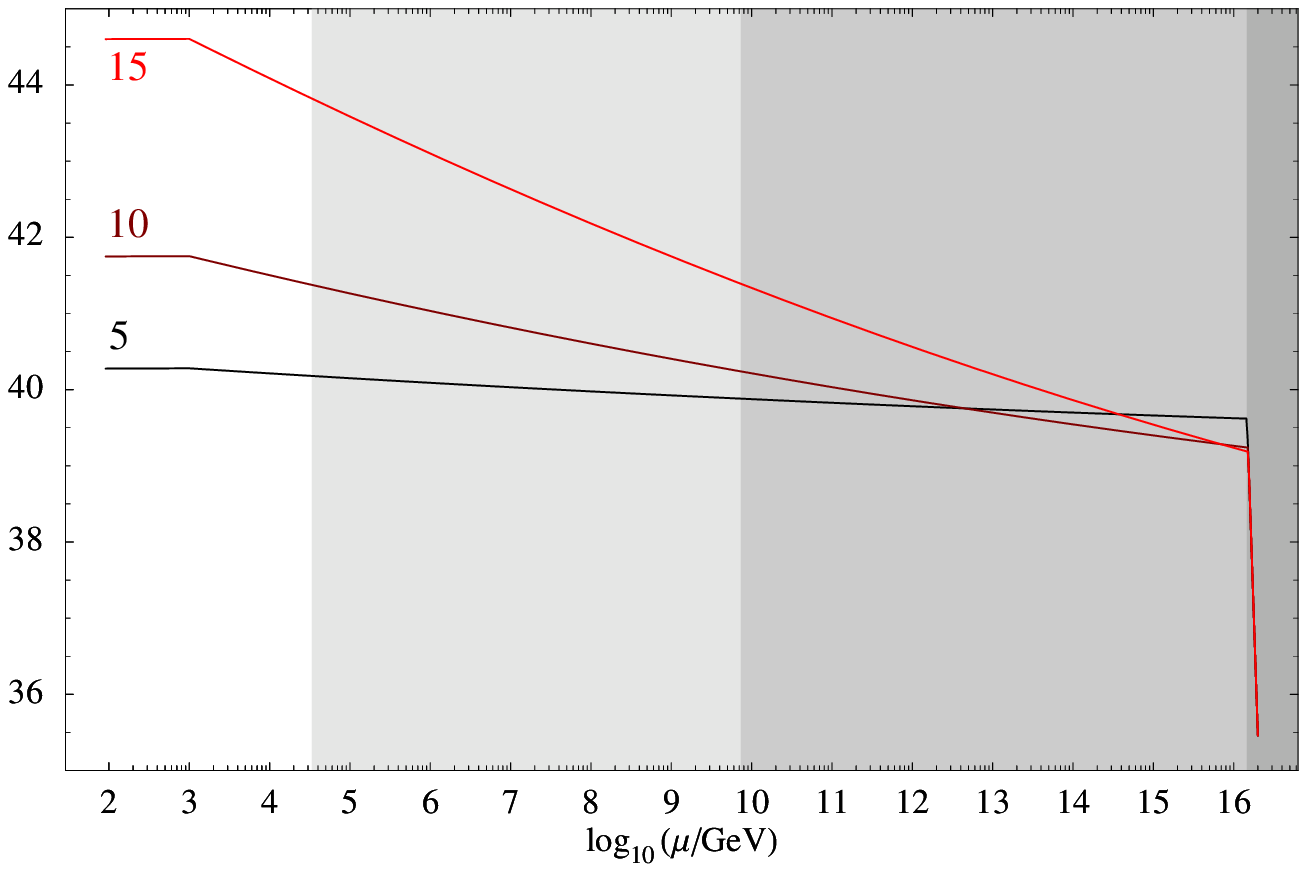}}}
\caption{Examples of running of $\theta_{12}$ in the case of MSSM and inverted mass
hierarchy. The dependence of $\theta_{12}$ on $\mu$ (a) for different
values of $m_1$, and    $\tan \beta = 10$,
(b) on $\tan \beta$ for $m_1 = 10^{-3}$ eV.
The value  $\varphi_2 =  0$ is taken.}
\label{fig:MSSMTheta12I}
\end{figure*}
%%%%%%%%%%%%%%%%%%%%%%%%%%%%%%%%%%%%%%%%%%%%%%%%%%%%%%%%%%%%%%%%%%%%%%%%%%%%%%%%%%

\subsection{MSSM and  inverted mass hierarchy}
%%%%%%%%%%%%%%%%%%%%%%%%%%%%%%%%%%%%%%%%%%%%%%%%%%%%%%%%

In the case of  inverted mass hierarchy,  the states $\nu_1$ and $\nu_2$
associated to 1-2 mixing are strongly degenerate.
Therefore, the RG effects are similar to those
in  the normal hierarchy case  for
$m_1 =  m_A \sim  5 \cdot 10^{-2}$ eV. So, the corrections,
$\Delta \theta_{12}$, are enhanced by the factor
\be
\frac{(\Delta \theta_{12})^{IH}}{(\Delta \theta_{12})^{NH}}
= \frac{(m_2^{IH})^2}{(m_2^{NH})^2},
\ee
where subscripts NH and IH stand for normal and inverted
mass hierarchy. This factor equals
\be
\frac{\Delta m_{13}^2}{\Delta m_{21}^2} ~~~{\rm or} ~~~~
\frac{(m_1^{IH})^2}{(m_1^{NH})^2}
\label{enhancm}
\ee
for the strong normal hierarchy
and normal ordering ($m_1 \approx m_2$) correspondingly.

In fig. \ref{fig:MSSMTheta12I} we show examples of
running of $\theta_{12}$ for different values of masses and phases. Dependences of  $\theta_{12}$ are well
described by $Q_{12}$,  as in the case of  normal mass hierarchy.
Notice that now the heaviest RH neutrino mass is determined
by $m_3$, and two others by $m_A$.
With increase of $m_3$ (now the lightest neutrino mass)
(fig. \ref{fig:MSSMTheta12Ia}) the range above
the seesaw scales, where the evolution of
$\theta_{12}$ is most strong,  increases. The change of $\theta_{12}$ below $M_3$
is slower being  of the same size for different values
of $m_3$  (until $m_3  \ll m_A$).
In this range the evolution  is essentially due to $Y_e$
couplings,  so that $\Delta\theta_{12} \propto \tan^2 \beta$
(fig. \ref{fig:MSSMTheta12Ib}).
The correction can be strongly suppressed for the opposite
CP-parities of $\nu_1$ and $\nu_2$:
$\Delta\theta_{12} \propto (1 + \cos \varphi)$.

%%%%%%%%%%%%ffff6%%%%%%%%%%%%%%%%%%%%%%%%%%%%%%%%%%%%%%%%%%%%%%%
%% \begin{figure}[H]
%% \centerline{
%% {\includegraphics[width=6.5cm]{MSSMTanBetaM1IContour}}\quad
%% }
%% \caption{Contours of constant RG corrections, $\Delta \theta_{12}$, in the
%% $\tan \beta - m_1$ plane in the case  of  MSSM and inverted  mass hierarchy
%% and  $\varphi_2 = 0$.
%% }
%% \label{fig:MSSMtanBetaM1I}
%% \end{figure}
%%%%%%%%%%%%%%%%%%%%%%%%%%%%%%%%%%%%%%%%%%%%%%%%%%%%%%%%%%%%%%%%%%%%

%%%%%%%%%%%%%%%%%%%%%%%%%%%%ffff7%%%%%%%%%%%%%%%%%%%%%%%%%%%%%%%%%%%%%%%%%
\begin{figure}%[H]
  \centerline{
{\includegraphics[width=8cm]{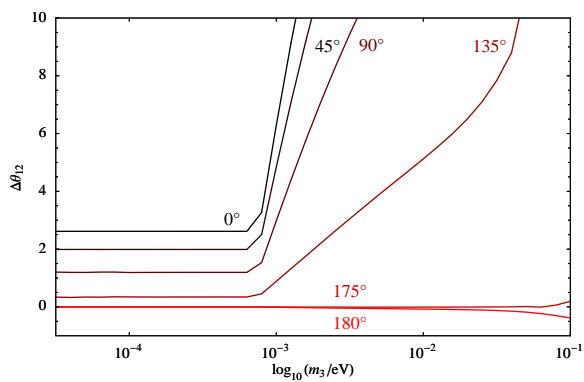}}\quad
}
\caption{The dependence of the RG correction $\Delta \theta_{12}$ on $m_1$
for different values of $\varphi_2$ (figures at the curves)
in the case of MSSM, the inverted  mass hierarchy and  $\tan\beta=10$.
%%(Values  of $\varphi_2$ increase from black to red.)
}
\label{fig:MSSMPhi2IL}
\end{figure}
%%%%%%%%%%%%%%%%%%%%%%%%%%%%%%%%%%%%%%%%%%%%%%%%%%%%%%%%%%%%%%%%%%%%%%%%%%%%%%%%%

\begin{figure}%[H]
\centerline{
{\includegraphics[width=6.5cm]{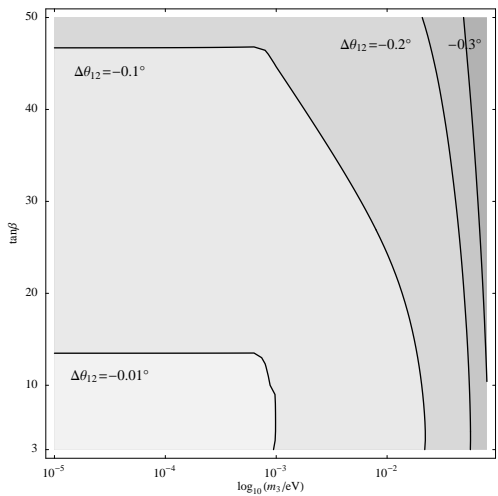}}\quad
}
\caption{Contours of constant RG corrections, $\Delta \theta_{12}$, in the
$\tan\beta - m_1$ plane in the case of  MSSM, inverted  mass hierarchy and
$\varphi = \varphi_2 = \pi$.
}
\label{Contour180}
\end{figure}

As in the case of normal hierarchy (ordering),
in a large part of the parameter space
the correction is positive, $\Delta\theta_{12} > 0$, due
to dominant effect of $P_{33}$.
For $\varphi_2 = 0$,
consistency of  the QLC prediction with data taken as
$\Delta\theta_{12} < 2^{\circ}$, implies
$\tan \beta < 10$ and $m_3 <  8 \cdot 10^{-4}$ eV.
For $\varphi_2 \sim  \pi$ corrections can be strongly suppressed, so that a
larger region of the parameter space becomes allowed.
The corrections become negative for $\varphi_2 = \pi$
(see figs.  \ref{fig:MSSMPhi2IL}, \ref{Contour180}) when the
leading RG effects are strongly suppressed and the running is mainly due to
sub-leading effect related to non-zero 1-3 mixing.
This possibility has been mentioned in \cite{qlc2}.
The sign of  contribution due to  non-zero $\theta_{13}$
to the RG running  of  $\theta_{12}$
due to  non-zero $\theta_{13}$
depends on the parameter (masses, phases) region.

%%%%%%%%%%%%%%%%%%ffff8%%%%%%%%%%%%%%%%%%%%%%%%%%%%%%%%%%%%%%%%%%%%
%%%%%%%%%%%%%%%%%%%%%%%%%%%%%%%%%%%%%%%%%%%%%%%%%%%%%%%%%%%%%%%%%%%%

In general, for non-zero $\theta_{13}$, the
contribution to $\Dot \theta_{12}$ is given by \cite{exactRGE}
\begin{widetext}
\begin{multline}
\frac{C_\nu \theta_{13}}{32\pi^2}\sin 2\theta_{23} \bigg[\left(\mathcal{Q}_{12}\cos
2\theta_{12}+
\mathcal{Q}_{13}s^2_{12}+\mathcal{Q}_{23}c^2_{12}\right)\cos\delta\\
+2\left(\frac{m_1 m_2}{\Delta m_{21}^2} \sin(\varphi_1-\varphi_2)+
\frac{m_1m_3}{\Delta m_{31}^2}\sin\varphi_1 s^2_{12}+
\frac{m_2m_3}{\Delta m^2_{32}}\sin\varphi_2 c^2_{12}\right)\sin\delta\bigg]\; .
\end{multline}
\end{widetext}
According to this equation
for $\varphi_2=180^\circ$, $\varphi_1=0^{\circ}$
and $\delta=180^\circ$, the dominant contribution
is determined by the combination
$-\frac{m_3+m_1}{m_3-m_1}\sin^2\theta_{12}\sin 2\theta_{23}$,
that is positive in the inverted hierarchy case, and
therefore $\theta_{12}$ decreases from high to low energies.

\section{RG effects in the Standard model}
%%%%%%%%%%%%%%%%%%%%%%%%%%%%%%%%%%%%%%%%%%%%%%%%%%%%%%%%%%%

In the standard model the evolution of $\theta_{12}$ is more complicated.
As we have already mentioned, apart from the Yukawa
coupling contributions described by eq.~(\ref{eqrun})
there are additional vertex diagrams that  cancel
in the SUSY case~\cite{Antusch:2002rr}. Furthermore, the vertex diagrams with
the gauge bosons become important: their contribution to
running between the seesaw scales influences
the flavor structure of mass matrix and therefore
changes $\theta_{12}$ \cite{exactRGE,manf13,aboveseesaw}.
The point is that individual RH neutrinos, $N_i$,  have
the flavor dependent couplings with the left handed components of  neutrinos.
Therefore the gauge boson corrections to the corresponding
couplings will influence the flavor structure.
Above the seesaw scales (where all RH neutrinos are operative)
and below the seesaw scales (where all RH neutrinos decouple),
flavor universality of the gauge interaction corrections is restored.
There is no simple analytic formula for the $\theta_{12}$ renormalization in the SM.

%%%%%%%%%%%%%%%%%%%%%%%%%%%%%%%%%ffff9ab%%%%%%%%%%%%%%%%%%%%%%%%%%%%%%%%%%%
\begin{figure*}
\centerline{
  \subfigure[~]{\label{fig:SMTheta12a}\includegraphics[width=7.5cm]{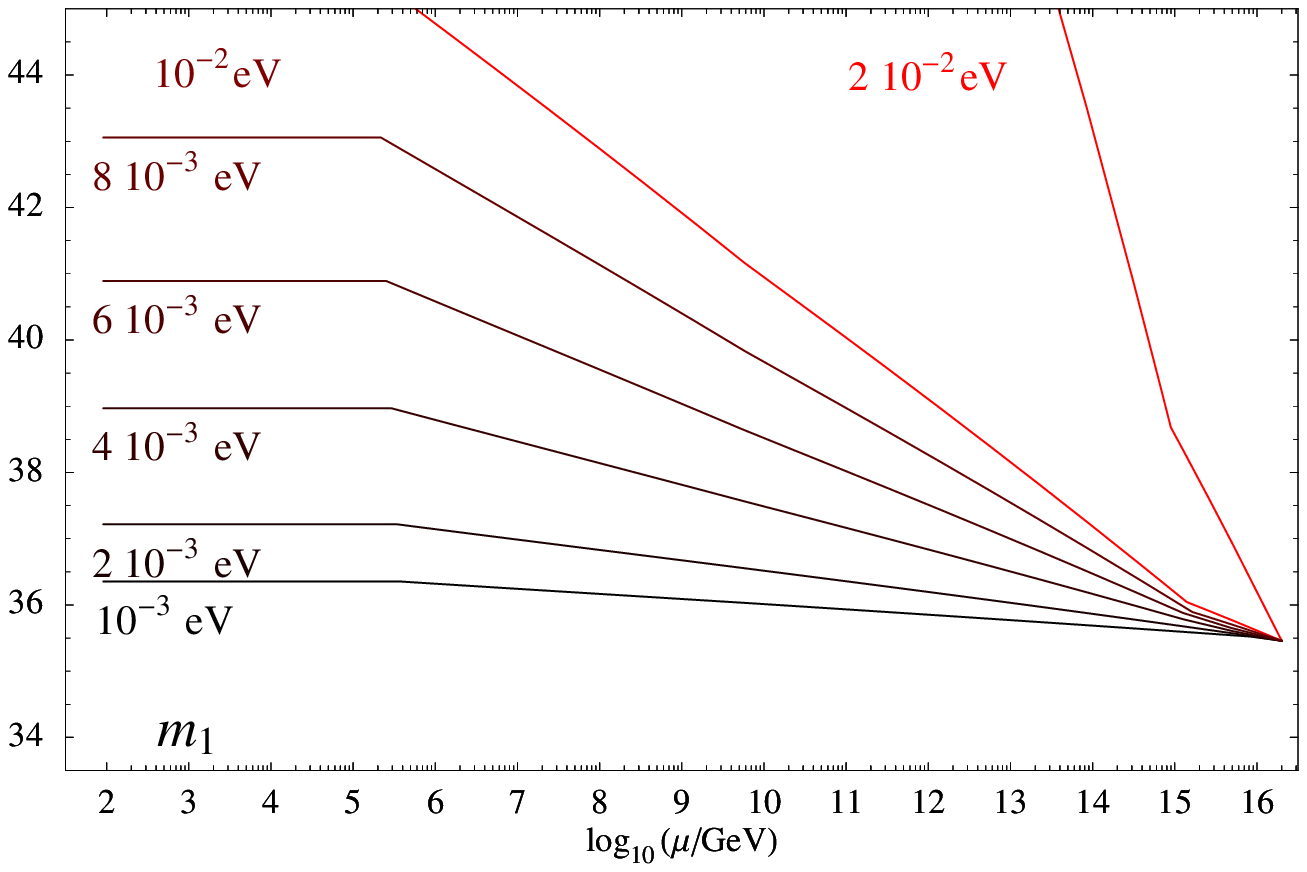}}\quad
  \subfigure[~]{\label{fig:SMTheta12b}\includegraphics[width=7.5cm]{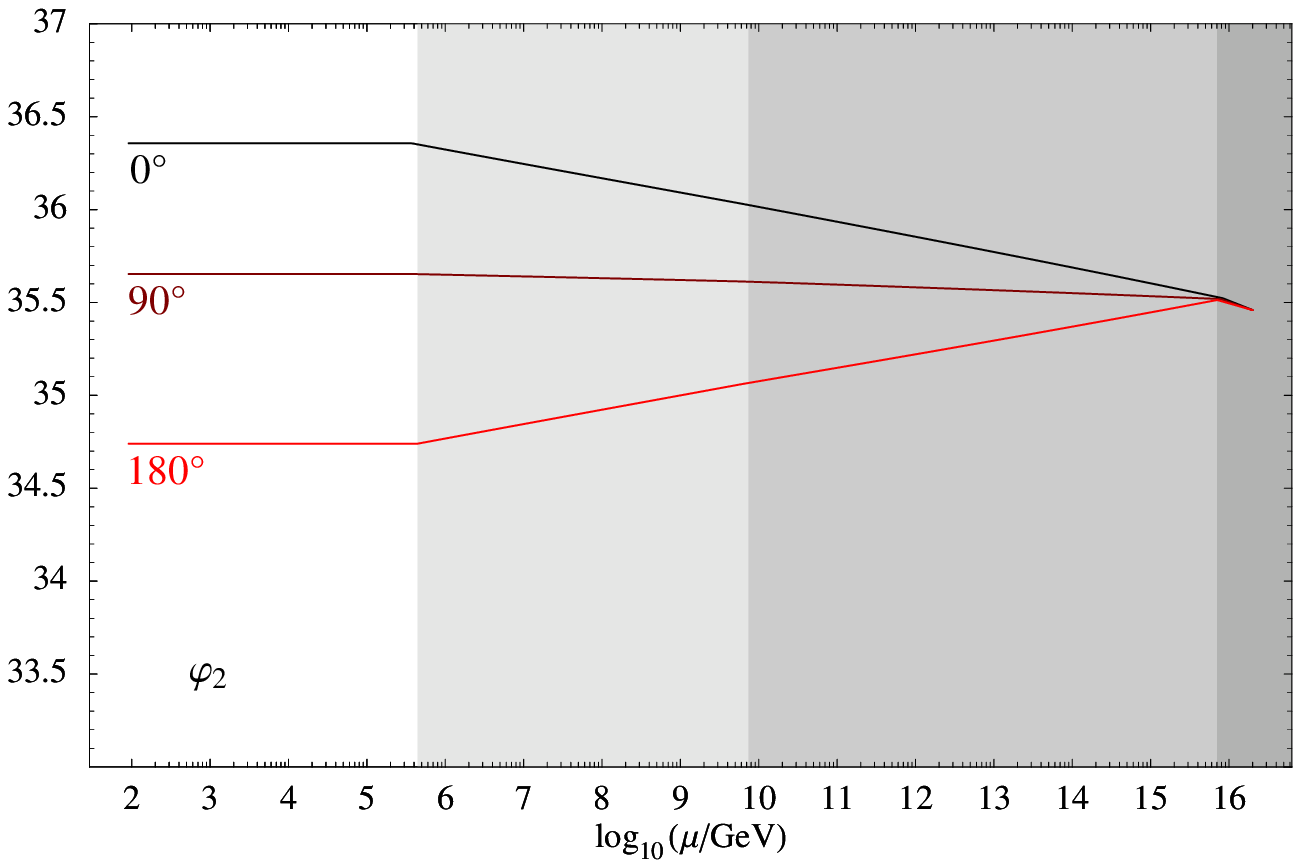}}}
\caption{Examples of running of $\theta_{12}$ in the case of SM and normal mass
hierarchy. The dependence of $\theta_{12}$ on $\mu$ (a) for different
values of $m_1$, and   $\varphi_2 = 0$,
(b) on $\varphi_2$ for $m_1 = 10^{-3}$ eV.
}
\label{fig:SMTheta12}
\end{figure*}
%%%%%%%%%%%%%%%%%%%%%%%%%%%%%%%%%%%%%%%%%%%%%%%%%%%%%%%%%%%%%%%%%%%%%%%%%%%%%%%%

In fig. \ref{fig:SMTheta12} we show examples of the $\theta_{12}$
scale evolution. Above the seesaw scales the running  is due to
the Yukawa interactions, $Y_{\nu}$, and the effect is well described by
the analytic results (\ref{eq:DotTheta12}). Below the seesaw scales, $\mu < M_1$,
the evolution is negligible: it is related
to $Y_e$ couplings that are small in the SM.
The main effect is collected  between the seesaw scales.
As we mentioned above, it is mainly
due to the gauge vertex corrections:
After decoupling of $N_3$ the D=5 operator
\be
\frac{1}{M_3} (Y_{\nu})_{ii} (U_{R})_{i3} (U_{R})_{3j} (Y_{\nu})_{jj}
L_i^T L_j H H
\ee
is formed.
The vertex diagram corrections to this operator due to the gauge (and also Yukawa)
interactions  produce running. Notice that for other RH neutrinos
that  do not decouple, the corresponding couplings
produce box diagrams with propagators of the RH neutrinos. Those diagrams
are finite and do not lead to logarithmic corrections.
The gauge interaction effect dominates since $N_3$ with the
largest Yukawa coupling is decoupled and $Y_e$ are small.
The corrections increase with $m_1$.

The most interesting dependence of $\Delta \theta_{12}$
is the one on the CP-violation phase
$\varphi_2$ (fig. \ref{fig:SMTheta12b}). The corrections are positive,
$\Delta \theta_{12} > 0$, for $\varphi_2 = 0$.
They are strongly suppressed for  $\varphi_2 \sim  \pi/2$,
in contrast to the SUSY case where suppression is realized for
$\varphi \sim \pi$. The corrections are negative
for  $\varphi_2 > \pi/2$.
The angle of zero corrections, $\varphi_2 (0)$, depends on
$m_1$ and in general deviates from $\pi/2$. The deviation
is due to the Yukawa interaction effects that produce the positive
shift for strong Yukawa coupling hierarchy
as we discussed before. The shift occurs both above
and between the seesaw scales (see fig. \ref{fig:SMTheta12b}).

\begin{figure}%[H]
\centerline{
{\includegraphics[width=7.5cm]{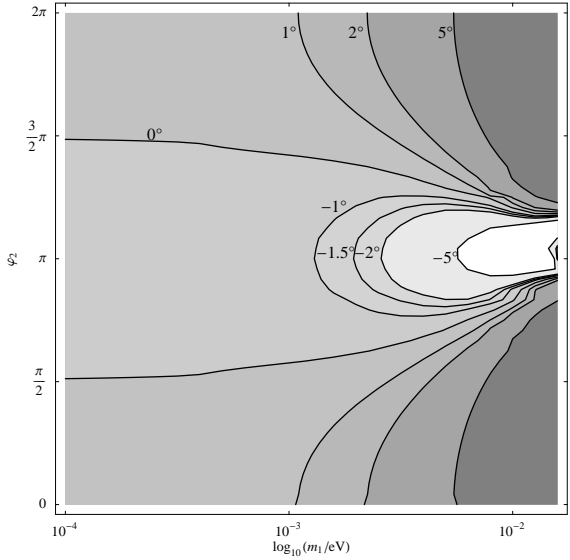}}
}
\caption{Contours of constant RG corrections to $\theta_{12}$ (figures at the curves)
in the $\varphi_2 - m_1$ plane in the case  of SM and normal mass hierarchy.
}
\label{fig:SMPhi2M1}
\end{figure}

\begin{figure}%[H]
\centerline{
{\includegraphics[width=8cm]{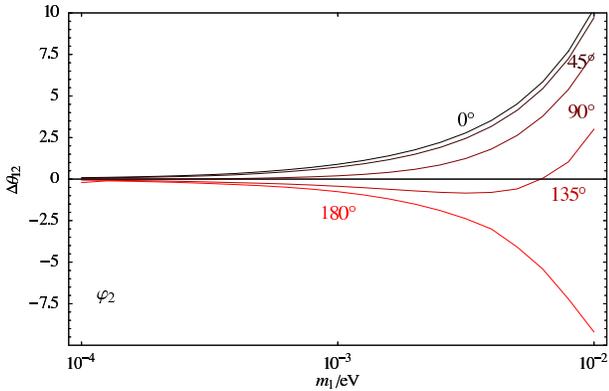}}\quad
  }
\caption{The dependence of the RG correction $\Delta \theta_{12}$ on $m_1$
for different values of $\varphi_2$ (figures at the curves) in the SM  with the normal mass
hierarchy.
}
\label{fig:SMPhi2L}
\end{figure}

In fig. \ref{fig:SMPhi2M1}  we show contours of
constant  corrections in the $m_1 - \varphi_2$ plane, and in fig. \ref{fig:SMPhi2L}
-- an explicit dependence of $\Delta \theta_{12}$ on $m_1$ for different values of
$\varphi_2$.
The line $\Delta \theta_{12} = 0$, is close to
$\varphi_2  =  \pi/2,~ 3\pi/2$ for  $m_1 \rightarrow 0$,  and it approaches
$\pi$ with increase of  $m_1$ when spectrum becomes strongly
degenerate. The pattern is nearly symmetric with respect to
$\varphi  = \pi$ for small $m_1$, the asymmetry
appears for  $m_1 > 3 \cdot 10^{-3}$ eV.

%%%%%%%%%%%%%%%%%%%%%%%%%ffff10%%%%%%%%%%%%%%%%%%%%%%%%%%%%%%%%%%%%%%%%%%%%%%%%%%%%%%%%%%
%%%%%%%%%%%%%%%%%%%%%%%%%%%%%%%%%%%%%%%%%%%%%%%%%%%%%%%%%%%%%%%%%%%%%%%%%%%

The line $\Delta \theta_{12} = 2^{\circ}$ restricts
the region consistent with the QLC relation.
Along the contours $\Delta \theta_{12} = -1.5^{\circ}$
the best fit experimental value for $\theta_{12}$
can be reproduced. This corresponds to
$m_1 > 2 \cdot 10^{-3}$ eV and $\varphi_2  \sim 5\pi/6 - 7\pi/6$.
Large negative corrections appear in the region
$m_1 > 5 \cdot 10^{-3}$ eV  and $\varphi_2  \sim \pi$.

%%%%%%%%%%%%%%%%%%%ffff9%%%%%%%%%%%%%%%%%%%%%%%%%%%%%%%%%%%%%%%%%%%%%%%%%%%%%%%%%
%%%%%%%%%%%%%%%%%%%%%%%%%%%%%%%%%%%%%%%%%%%%%%%%%%%%%%%%%

\section{Renormalization of 13 mixing. Level crossing. Evolution above $M_{GUT}$}
%%%%%%%%%%%%%%%%%%%%%%%%%%%%%%%%%%%%%%%%%%%%%%%%%%%%%%%%%%%%%%

\subsection{Renormalization of 13 mixing}
%%%%%%%%%%%%%%%%%%%%%%%%%%%%%%%%%%%%%%%%%%%

In the scenario discussed in this paper, the 1-3 mixing is non-zero and relatively large
at the  boundary  (\ref{pred13}).
%Furthermore, at tree level  the mixing is not small being
%close to the present $1\sigma$  bound.
Notice that  $\theta_{13}$ (i) interferes with
1-2 mixing in the QLC relation as we discussed before;
(ii) produces sub-leading effects in renormalization of $\theta_{12}$,
(iii) can provide further bounds on the
considered scenario if  RG corrections are  positive and large.

The dominant contribution to the renormalization of  $\theta_{13}$
is given by \cite{exactRGE,manf13,aboveseesaw}
\be
64\pi^2\Dot\theta_{13}=C_\nu\sin2\theta_{12}\sin2\theta_{23} (\mathcal{A}_{13}-\mathcal{A}_{23}),
\label{evol13}
\ee
where
\be
\mathcal{A}_{i3} \equiv \frac{1}{\Delta m^2_{3i}}
[(m_i^2 + m_3^2)\cos\delta + 2m_i m_3 \cos(\delta-\varphi_i)]\;.
\ee
In our case  $\sin2\theta_{12} > 0$,
$\sin2\theta_{23} > 0$, $\delta\approx 180^\circ$
and  for vanishing Majorana CP phases, $\varphi_i = 0$,
the dominant contribution can be approximated to
\be
64\pi^2\Dot\theta_{13}=
C_\nu\sin2\theta_{12}\sin2\theta_{23} (\mathcal{Q}_{23}-\mathcal{Q}_{13}),
\label{evol13b}
\ee
and the last factor in
(\ref{evol13b}):  $(\mathcal{Q}_{23} - \mathcal{Q}_{13})=
(\mathcal{A}_{13} - \mathcal{A}_{23})$ is negative, irrespective of
the mass hierarchy.  Consequently $\theta_{13}$ increases when running to low energies.
For  non-vanishing phases $\varphi_i$  this factor can be positive, thus leading to a decrease
of
$\theta_{13}$ when  $\mu$ decreases.
%Yes, but it also depends on the phases.

In the case of strong mass hierarchy eq.~(\ref{evol13}) gives
\be
64\pi^2\Dot\theta_{13} = -2 \sin2\theta_{12}\sin2\theta_{23}\cos(\delta-\varphi_2)
\sqrt{\frac{\Delta m^2_{21}}{\Delta m^2_{31}}}.
\label{evol13h}
\ee
The running is suppressed by small mass ratio.
Therefore only a  small RG effect on 1-3 mixing appears for the hierarchical
(normal as  well as inverted) case.
For instance, we find that for the parameter sets used in figure 2 (MSSM),
the correction $\Delta \theta_{13}$ is always smaller than $0.2^{\circ}$.
In the SM, it is  smaller than $0.3^{\circ}$.

For the degenerate spectrum, there can be a larger effect
which strongly depends on the CP-phases.
From (\ref{evol13}) we find
\be
64\pi^2\Dot\theta_{13} \approx 2\sin2\theta_{12}\sin2\theta_{23} \frac{m_1^2}{\Delta m_{31}^2}
[\cos(\delta-\varphi_1) - \cos(\delta-\varphi_2)].
\label{evol13d}
\ee
Notice that for zero CP phases  the cancellation occurs again.
In  the MSSM  for  $m_1=0.03$ eV and $\tan\beta=50$,
we find $\Delta \theta_{13} \sim 0.5^{\circ}$.
In contrast, for $\delta = \varphi_1 = \pi$ and $\varphi_2 = 0$
the two terms in (\ref{evol13d}) sum up and we obtain
positive running:
$64\pi^2\Dot\theta_{13} \approx 4\sin2\theta_{12}\sin2\theta_{23}$.
Consequently $\theta_{13}$ becomes smaller at low energies.

%%%%%%%%%%%%%%%%%%%%%%%%%%%%%%%%%%%%%%%%%%%%%%%%%%
\subsection{Level crossing points}
%%%%%%%%%%%%%%%%%%%%%%%%%%%%%%%%%%%%%%%%%%%%%%%%%%

As we have established in sect.~3 the spectrum of the right--handed
Majorana neutrinos is generically hierarchical. However, there are the  level
crossing points, where two of the RH neutrino masses become
equal \cite{AFS}.
The case of degeneracy of two lighter RH neutrino states,
$M_1 \approx M_2$,   is of special interest from the point of view of
generation of the baryon asymmetry in the Universe.
In this case the resonance leptogenesis becomes possible
which produces large enough asymmetry in spite of smallness
of the masses and  consequently, large wash out effect.

From (\ref{123mb}) we find
\be
M_1 = \frac{2 m_t^2 \epsilon'^4}{m_1 + m_2}, ~~~
M_2 =
\frac{2 m_t^2 \epsilon^2(m_1 + m_2)}{(m_1 + m_2)m_3 + 2 m_1 m_2}.
\ee
(Here the Majorana phases are included in $m_i$).
It is easy to see that  due to  smallness
of $\epsilon$ the condition  $M_1  \approx M_2$ can be satisfied
only in the case of strong mass degeneracy
$|m_1| \approx |m_2| \approx m_0$
when
\be
m_1 + m_2  = \frac{\Delta m_{21}^2}{2m_0} \approx 0.
\label{condition}
\ee
Then from the condition $M_1  \approx M_2$ we find
\be
m_0 = \sqrt{\frac{\Delta m_{21}^2}{2\sqrt{2}\epsilon}} \sim 0.1~
{\rm eV}.
\ee
In this special case the mass
\be
M_1 \approx M_2 \frac{4 m_t^2 \epsilon'^4 m_0}{\Delta m_{21}^2}
=  M_1^{NH} \frac{2m_0}{\sqrt{\Delta m_{21}^2}}
\ee
is enhanced by factor $2m_0/\sqrt{\Delta m_{21}^2} \sim 20$
and the third mass is much smaller than in the hierarchical case:
\be
M_3 \approx \frac{m_t^2}{2 m_3},
\ee
that is, smaller by factor $m_1^{NH}/m_3 < 10^{-3}$.

The level crossing condition  (\ref{condition}) implies the opposite
CP-violating phases; it coincides with the condition of
strong suppression of  the RG effects.
It also  implies  smallness of the 11-element of
$m_{bm}$ matrix. The condition for level crossing differs from that in
\cite{AFS} since here we require the neutrino Dirac  matrix to be
diagonal in the basis where the mass matrix
of light neutrinos has exactly  bimaximal form.
If instead we use a generic matrix with non-maximal 1-2 mixing
the level crossing condition can be realized for the hierarchical
spectrum \cite{AFS}.

%%%%%%%%ffff12%%%%%%%%%%%%%%%%%%%%%%%%%%%%%%%%%%%%%%%%%%%%%%%%%%%%%%%%%%%%%%%%%%
\begin{figure}%[H]
\centerline{
{\includegraphics[width=7.5cm]{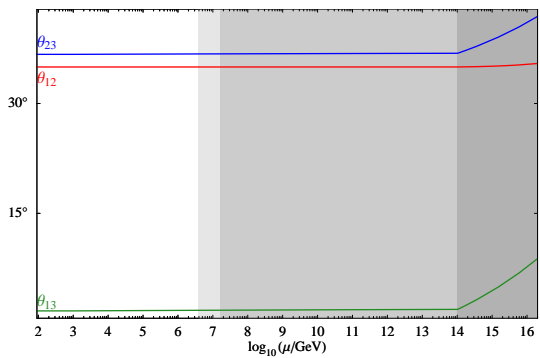}}}
\caption{Examples of running of mixing angles in the case
of $M_1  \approx M_2$ in  MSSM
and normal mass ordering.
We show the dependence of $\theta_{12}$, $\theta_{13}$ $\theta_{23}$
on $\mu$ for  $\tan \beta= 10$,   $\varphi_1=0$, $\varphi_2 = \pi$ and  $m_1=0.13$ eV.
}
\label{LevelC}
\end{figure}
%%%%%%%%%%%%%%%%%%%%%%%%%%%%%%%%%%%%%%%%%%%%%%%%%%%%%%%%%%%%%%%%%%%%%%%%%%%%%%%%%%

In fig. \ref{LevelC} we show scale evolution of the mixing angles
for parameters that correspond to the level crossing point
$M_1  = M_2$. In this point $M_1 = M_2 =  8\cdot 10^6$ GeV,
$M_3 = 8\cdot 10^{13}$ GeV, $\varphi_1=0$, $\varphi_2 = \pi$,  $m_1=0.13$ eV.
The angle  $\theta_{12}$ evolves very weakly due to
cancellation $Q_{12} = S_{12} \approx 0$ related to (\ref{condition}).
In contrast,
the 1-3 mixing evolves substantially above thresholds:
$\Delta \theta_{13} = 7^{\circ}$.
%The resulting  value $\theta_{13}$ is experimentally excluded.
The 2-3 mixing shows relatively weak evolution, that, however,
can  influence the second QLC relation.

We find  that in this crossing point  the solar mass squared
difference becomes large even if it is  very small  at the boundary.
So,  the 1-2 split has the radiative origin.
The 1-3 split decreases by factor $\sim 2$.

\subsection{Evolution above the GUT scale}
%%%%%%%%%%%%%%%%%%%%%%%%%%%%%%%%%%%%%%%%%%%%%%%%%%%%%%%%%%%%%%%%%

For $M_F > M_{GUT}$  one should perform running also above the GUT scale.
Restoration of the GUT symmetry and unification of the
gauge couplings does not  prevent from different running of the
Yukawa couplings, and therefore, from change of mixing angles.
Renormalization of mixing angles would stop after
possible unification of the Yukawa couplings  which can be related, {\it e.g.}, to
restoration at $M_F$ a non-Abelian flavor symmetry. An alternative is the
boundary at the string or Planck scale, where the Yukawa couplings are formed
and  their properties are determined immediately by some symmetry
or/and string selection rules.

\begin{figure}%[H]
  \centerline{
{\includegraphics[width=8cm]{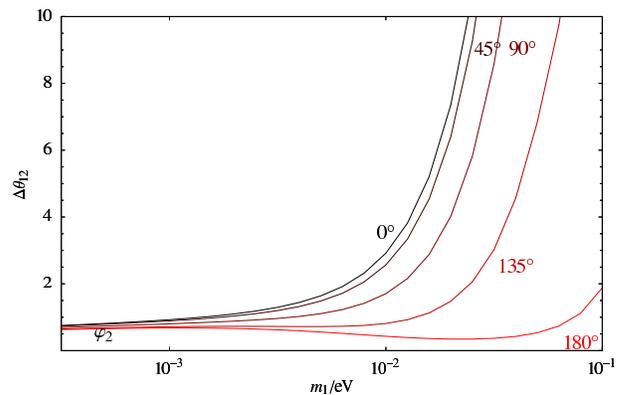}}\quad
}
\caption{The dependence of the RG correction $\Delta \theta_{12}$ on $m_1$
for different values of $\varphi_2$ (figures at the curves)
in the MSSM  with the normal mass
hierarchy and  $\tan\beta=10$.   The boundary condition is at $M_{Pl}$.
}
\label{fig:MSSMPhi2LPl}
\end{figure}

For illustration we performed the running in the MSSM up to the
Planck scale (ignoring possible GUT effects).
In fig. \ref{fig:MSSMPhi2LPl}  we show  the dependence of $\Delta \theta_{12}$
on $m_1$ for the  same (QLC) initial conditions at the
Planck scale:  $M_F = M_{Pl} = 1.2 \cdot  10^{19}$ GeV.
The RG effect becomes much larger.
In particular the contribution from the region above the seesaw scale due to
large Yukawa coupling $Y_{\nu}$ increases substantially.
It is enhanced  in comparison with the case of  running up to $M_{GUT }$
by the factor
\be
\frac{\log(M_{Pl}/M_3)}{\log(M_{GUT}/M_3)}
\ee
that can be as large as  3 - 5 in some cases.
Still for $\varphi_2 = \pi$ or for small $m_1$ the RG effects are
suppressed and can be consistent with the QLC relations.

Similar RG effects are expected in the SU(5) model with the
RH singlet neutrinos. In fact, no new diagrams with large
$Y_{\nu}$ appear.  Effect of the charged lepton couplings $Y_e$ is  enhanced
by factor 4 above $M_{GUT}$
due to the loop diagrams with down quarks (squarks) and $H^{1/3}$ charged
Higgs bosons (Higgsinos).

%%%%%%%%%%%%%%%%%%ffff11%%%%%%%%%%%%%%%%%%%%%%%%%%%%%%%%%%%%%%%%%%%%%%%%%%%%
%%%%%%%%%%%%%%%%%%%%%%%%%%%%%%%%%%%%%%%%%%%%%%%%%%%%%%%%%%%%%%%

The flavor-diagonal parts of the RG equations do
influence the angles only indirectly  through the change
of the mass eigenvalues. Thus, the  main effect of these interactions
is due to the evolution of $\Delta m_{12}^2$.

\section{Conclusion}
%%%%%%%%%%%%%%%%%%%%%%%%%%%%%%%%%%%%%%%%%%%%%%%%%%%%%%%%%%%%%%%%%

Experimental results on the 1-2 and 2-3 mixing in the quark and lepton sectors
show certain correlations that  can be interpreted as the quark-lepton complementarity.
We considered the  QLC scenario
in which the bimaximal mixing
follows from  diagonalization of the neutrino mass matrix.
In the lowest order of the perturbation theory,
the value of angle $\theta_{12}$ predicted in this scenario
is about $\sim 1\sigma$ larger than the best fit experimental value.
It coincides practically with the value given by the tri-bimaximal mixing.
We commented on  implications of this equality as well as on
perspectives of future tests  of the QLC relations.

In this paper, we assumed that the QLC relations are not accidental
coincidences, but consequence of the quark-lepton symmetry and
additional structure in the theory that produces the bimaximal mixing.
Here the accidental coincidences would mean that  values of
mixing angles  are the result of  interplay of two or more independent
contributions.
In this connection, we proposed a  realization of the QLC scenario
that provides the closest relation between quarks and leptons. It is based on

- the seesaw type-I mechanism  that generates
the bimaximal mixing due to specific
structure of the RH neutrino mass matrix;

- approximate equalities  of the Dirac mass matrices:
$m_u \approx m_D$, $m_l \approx  m_d$ that follow from the
approximate quark-lepton
symmetry or unification. A certain small violation of equalities of these matrices
produces difference of mass hierarchies  but does not affect
substantially the mixing.

The only additional (and, in fact, unavoidable) factor that can
affect  the QLC relations is RG corrections.

One of the consequences  of the proposed scenario  is a very strong hierarchy
of the RH neutrino masses, apart from several particular level crossing points.
The latter are realized for strongly degenerate spectrum of light neutrinos,
and particular values of the CP-violating phases.
This determines  substantially the RG effects.

We performed a systematic study of the RG effects in the  SM
and MSSM.
%Let us summarize our main results.
We find that in the  MSSM, the RG corrections to $\theta_{12}$
are  generically positive due to a
dominant effect of the Yukawa coupling $Y_{33}$.
So, these corrections worsen agreement of the predicted $\theta_{12}$
with data.

In the MSSM small negative corrections, $|\Delta \theta_{12}| < 0.5^{\circ}$,
can appear for the opposite CP parities of $\nu_1$ and $\nu_2$
and inverted mass hierarchy, in which case the  main terms in the RG equations
are strongly suppressed and
running is due to the sub-leading effects related to the non-zero 1-3 mixing.
The correction increases with $m_1$ and strongly
depends on the relative Majorana phase. For
$\varphi_2 = 0$ the consistency of the  QLC prediction
for $\theta_{12}$  with data implies strong mass  hierarchy of light
neutrinos  and small
$\tan \beta$. For $\varphi_2 = \pi$ corrections
are suppressed and even the degenerate spectrum
becomes allowed.
For the inverted mass hierarchy the corrections are generically
enhanced  by larger values of masses of $\nu_1$ and $\nu_2$.

The situation is qualitatively different in the SM. Here
important contributions follow from the  vertex corrections to the D=5
operator in the range between the seesaw scales.
The Yukawa couplings (especially for small $m_1$) give
sub-leading contribution.
The RG corrections are negative in the interval $\varphi_2 = \pi/2  - 3\pi/2$
for small $m_1$ and the range of negative corrections  becomes narrower,
$\varphi_2 = (0.9 - 1.2) \pi$,
for the degenerate neutrinos $\nu_1$ and $\nu_2$.

Corrections depend substantially on the boundary scale $M_F$.
The value $\Delta \theta_{12}$ can be enhanced by factor 2-5 if $M_F$
increases from $M_{GUT}$ to $M_{Pl}$.

For the hierarchical mass spectrum  renormalization of
the 1-3 mixings is, in general, small: $\Delta \theta_{13} \sim 0.2^{\circ} - 0.3^{\circ}$.
The correction can be large, $\Delta \theta_{13} \sim \theta_{13}$, for the  degenerate spectrum.
The sign of correction depends on values of CP-violating phases. \\

In conclusion, in a large part of the parameter space
especially for the strong mass hierarchy
and  opposite CP phases of $\nu_1$ and $\nu_2$,
the RG corrections to the QLC relation are
small.  The corrections are positive in the MSSM
apart from small region of parameter space that  corresponds to the
degenerate spectrum of light neutrinos and their opposite CP parities.
The corrections  are negative in the SM for $\varphi_2 > \pi/2$.
In the considered QLC scenario the RG corrections  allow one to reproduce
the best fit experimental value of $\theta_{12}$ exactly.

\begin{acknowledgments}
%\section*{Acknowledgements}
The work  of A.~Yu.~S.  has been supported in part by the Alexander von Humboldt Foundation. M.~A.~S. acknowledges support from the ``Deutsche Forschungsgemeinschaft'' in the
``Sonderforschungsbereich 375 f\"ur Astroteilchenphysik'' and under project number RO--2516/3--1.
\end{acknowledgments}


\begin{thebibliography}{99}

\bibitem{mohsmi}For recent review see R.~N.~Mohapatra and A.~Y.~Smirnov,
  %``Neutrino mass and new physics,''
  hep-ph/0603118.

\bibitem{qlc}A. Yu. Smirnov, hep-ph/0402264.

\bibitem{qlc1}
M. Raidal, {\it Phys. Rev. Lett.}  {\bf 93}, 161801 (2004).

\bibitem{qlc2}
H. Minakata, A. Yu. Smirnov, {\it Phys. Rev.} D {\bf 70},
073009 (2004).

\bibitem{petcov}
S.~T.~Petcov and A.~Y.~Smirnov,
Phys.\ Lett.\ B {\bf 322}, 109 (1994).


\bibitem{bm}F. Vissani, hep-ph/9708483;
V.~Barger, S.~Pakvasa, T.~Weiler and K.~Whisnant, Phys.\ Lett.\ B
{\bf 437}, 107 (1998); A.~Baltz, A.S.~Goldhaber and M.~Goldhaber,
Phys.\ Rev.\ Lett. {\bf 81} 5730 (1998); G.~Altarelli and
F.~Feruglio, Phys.\ Lett.\ B {\bf 439}, 112 (1998); M.~Jezabek and
Y.~Sumino, Phys.\ Lett.\ B {\bf 440}, 327 (1998); D. V. Ahluwalia,
Mod. Phys. Lett. {\bf A13}, 2249 (1998).



\bibitem{parametr}
M.~Jezabek and Y.~Sumino,
%``Neutrino masses and bimaximal mixing,''
Phys.\ Lett.\ B {\bf 457}, 139 (1999);
C.~Giunti and M.~Tanimoto, Phys.\ Rev.\ D {\bf 66}, 053013 (2002);
W.~Rodejohann, Phys.\ Rev.\ D {\bf 69}, 033005 (2004);
P.~H.~Frampton, S.~T.~Petcov and W.~Rodejohann,
Nucl.\ Phys.\ B {\bf 687}, 31 (2004);

\bibitem{qlc-fm} P. Frampton and R. N. Mohapatra, JHEP {\bf 0501}, 025
(2005).



\bibitem{shift}
J. Ferrandis and S. Pakvasa, Phys.\ Rev.\ D {\bf 71}, 033004 (2005);
D.~Falcone, hep-ph/0509028; A.~Ghosal and D.~Majumdar, hep-ph/0505173;
F.~Gonzalez Canales and A.~Mondragon,
hep-ph/0606175.


\bibitem{qlc-km}
S. Antusch, S. F. King and R. N. Mohapatra,
Phys.\ Lett.\ B {\bf 618}, 150 (2005).



\bibitem{qlc-cp} T.~Ohlsson and G.~Seidl,  Nucl.\ Phys.\ B {\bf 643}, 247
(2002);
S.~Antusch and S.~F.~King, Phys.\ Lett.\ B {\bf 631}, 42 (2005);
I.~Masina,  Phys.\ Lett.\ B {\bf 633}, 134 (2006); J.~Harada,
hep-ph/0512294;
  B.~C.~Chauhan, M.~Picariello, J.~Pulido and E.~Torrente-Lujan,
  hep-ph/0605032.

\bibitem{Hochmuth}
  K.~A.~Hochmuth and W.~Rodejohann,
  %``Low And High Energy Phenomenology Of Quark-Lepton Complementary
  %Scenarios,''
  hep-ph/0607103.

\bibitem{qlc-ren}
Sin Kyu Kang, C. S. Kim, Jake Lee,
Phys.\ Lett.\ B {\bf 619}, 129 (2005);
K.~Cheung, S.~K.~Kang, C.~S.~Kim and J.~Lee,
Phys.\ Rev.\ D {\bf 72}, 036003 (2005);
A.~Dighe, S.~Goswami and P.~Roy, hep-ph/0602062.



\bibitem{parametr2}
N.~Li and B.~Q.~Ma, Phys.\ Lett.\ B {\bf 600}, 248 (2004);
N.~Li and B.~Q.~Ma, Eur.\ Phys.\ J.\ C {\bf 42}, 17 (2005).
N.~Li and B.~Q.~Ma,
%``Unified parameterization of quark and lepton mixing matrices,''
Phys.\ Rev.\ D {\bf 71}, 097301 (2005);
Z.~z.~Xing, Phys.\ Lett.\ B {\bf 618}, 141 (2005).


\bibitem{par-gen}
A.~Datta, L.~Everett  and P.~Ramond, Phys.\ Lett.\ B {\bf 620}, 42 (2005);
T.~Ohlsson,  Phys.\ Lett.\ B {\bf 622}, 159 (2005);
L.~L.~Everett, Phys.\ Rev.\ D {\bf 73}, 013011 (2006).


\bibitem{sees} P. Minkowski, {\it Phys. Lett.} B {\bf 67} 421 (1977);
T. Yanagida, in {\it Proc. of Workshop on Unified Theory and Baryon
number in the Universe}, eds. O. Sawada and A. Sugamoto, KEK, Tsukuba, (1979);
M. Gell-Mann, P. Ramond and R. Slansky,  in {\it Supergravity}, eds P.
van Niewenhuizen and
D. Z. Freedman (North Holland, Amsterdam 1980);
P. Ramond, {\it  Sanibel talk}, retroprinted as hep-ph/9809459;
S. L. Glashow, in {\it Quarks and Leptons}, Carg\`ese lectures, eds M. L\'evy,
(Plenum, 1980, New York) p. 707;
R. N. Mohapatra and G. Senjanovi\'c, {\it Phys. Rev. Lett.} {\bf 44}, 912 (1980).


\bibitem{GST}R. Gatto, G. Sartori and M. Tonin, Phys. Lett. B{\bf 28},
128 (1968).

\bibitem{jarlskog}
C.~Jarlskog, Phys.\ Lett.\ B {\bf 625}, 63 (2005).

\bibitem{sno}SNO Collaboration (B. Aharmim et al.). {\it Phys. Rev.} C
{\bf 72}, 055502 (2005).

\bibitem{sv}A. Strumia, F. Vissani, {\it Nucl. Phys.} B {\bf 726}, 294 (2005).

\bibitem{bari}G. L. Fogli et al,  hep-ph/0506083.

\bibitem{12future}
  A.~Bandyopadhyay, S.~Choubey, S.~Goswami and S.~T.~Petcov,
  Phys.\ Rev.\ D {\bf 72} (2005) 033013;
  J.~F.~Kopp, M.~Lindner, A.~Merle and M.~Rolinec, hep-ph/0606151.

\bibitem{tbm}
L. Wolfenstein,  {\it Phys. Rev.} D {\bf 18}, 958 (1978);
P. F. Harrison, D. H. Perkins and W. G. Scott,
{\it Phys. Lett.} B {\bf 458}, 79 (1999),
{\it Phys. Lett.} B {\bf 530}, 167 (2002).



\bibitem{AFS}
  E.~K.~Akhmedov, M.~Frigerio and A.~Y.~Smirnov,
  %``Probing The Seesaw Mechanism With Neutrino Data And Leptogenesis,''
  JHEP {\bf 0309} (2003) 021.




%\bibitem{so10}H. Georgi, {\it In Coral Gables 1979 Proceeding, Theory and experiment
%in high energy physics}, New York 1975, 329 and H. Fritzsch and P. Minkowski, Annals
%Phys. {\bf 93} 193 (1975).

\bibitem{dss}R. N. Mohapatra, {\it Phys. Rev. Lett.}  {\bf 56}, 561 (1986);
R. N. Mohapatra and J. W. F. Valle, {\it Phys. Rev.} D {\bf 34}, 1642 (1986).


\bibitem{scre}M. Lindner, M. A. Schmidt, A. Yu. Smirnov, {\it JHEP} {\bf 0507},
048 (2005).

\bibitem{kim}O. Vives, hep-ph/0504079; J. E. Kim and J. C. Park, hep-ph/0512130.

\bibitem{pati}J. C. Pati and A. Salam, {\it Phys. Rev.} D {\bf 10}, 275 (1974).

\bibitem{rgeMat} P.~H.~Chankowski and Z.~Pluciennik, {\it Phys.\ Lett.\ B} {\bf 316}, 312 (1993);
  K.~S.~Babu, C.~N.~Leung and J.~T.~Pantaleone, {\it Phys.\ Lett.\ B} {\bf 319}, 191 (1993).

\bibitem{rge-eq}
J. A. Casas {\it et al.,} {\it Nucl. Phys.} B {\bf 573}, 652 (2000);
S. Antusch, M. Drees, J. Kersten, M. Lindner, M. Ratz, Phys.
Lett. B {\bf 519}, 238 (2001);  P.~H.~Chankowski and S.~Pokorski,
  Int.\ J.\ Mod.\ Phys.\ A {\bf 17}, 575 (2002);
  P.~H.~Chankowski, W.~Krolikowski and S.~Pokorski,
  Phys.\ Lett.\ B {\bf 473}, 109 (2000)

\bibitem{massmatrrg}
M.~K.~Parida, C.~R.~Das and G.~Rajasekaran,
  %``Radiative stability of neutrino-mass textures,''
  Pramana {\bf 62}, 647 (2004);
C.~Hagedorn, J.~Kersten and M.~Lindner,
  %``Stability of texture zeros under radiative corrections in see-saw
  %models,''
  Phys.\ Lett.\ B {\bf 597}, 63 (2004);
  T.~Miura, E.~Takasugi and M.~Yoshimura,
  %``Quantum effects for the neutrino mixing matrix in the democratic-type
  %model,''
  Prog.\ Theor.\ Phys.\  {\bf 104}, 1173 (2000),
M.~Frigerio and A.~Y.~Smirnov,  JHEP {\bf 0302}, 004 (2003).

\bibitem{manfth}
M. Tanimoto, Phys. Lett. B{\bf 360}, 41 (1995);
S.~F.~King and N.~N.~Singh, Nucl.\ Phys.\ B {\bf 591}, 3 (2000).

%%%%%%%%%%%%%%%%general corr analyt%%%%%%%%%%%%%%%%%%%%%%%%%%%%%%%
\bibitem{exactRGE}
  S.~Antusch, J.~Kersten, M.~Lindner and M.~Ratz,
  %``Running neutrino masses, mixings and CP phases: Analytical results and
  %phenomenological consequences,''
  Nucl.\ Phys.\ B {\bf 674}, 401 (2003).
\bibitem{manf13}
J.~w.~Mei and Z.~z.~Xing,
  %``Radiative generation of Theta(13) with the seesaw threshold effect,''
  Phys.\ Rev.\ D {\bf 70}, 053002 (2004);
  S.~Antusch, P.~Huber, J.~Kersten, T.~Schwetz and W.~Winter,
  %``Is there maximal mixing in the lepton sector?,''
  Phys.\ Rev.\ D {\bf 70}, 097302 (2004).

\bibitem{aboveseesaw}
  S.~Antusch, J.~Kersten, M.~Lindner, M.~Ratz and M.~A.~Schmidt,
  JHEP {\bf 0503}, 024 (2005).

%\cite{Antusch:2002rr}

\bibitem{bmrad}
  S.~Antusch, J.~Kersten, M.~Lindner and M.~Ratz,
  %``The LMA solution from bimaximal lepton mixing at the GUT scale by
  %renormalization group running,''
  Phys.\ Lett.\ B {\bf 544}, 1 (2002);
 T.~Miura, T.~Shindou and E.~Takasugi,
  %``The renormalization group effect to the bi-maximal mixing,''
  Phys.\ Rev.\ D {\bf 68}, 093009 (2003);
T.~Shindou and E.~Takasugi,
  %``The role of Majorana CP phases in the bi-maximal mixing scheme:
  %Hierarchical Dirac mass case,''
  Phys.\ Rev.\ D {\bf 70}, 013005 (2004).


\bibitem{degen}
J. A. Casas {\it et al.,} {\it Nucl. Phys.} B {\bf 556}, 3 (1999),
{\it ibidem},  {\bf 569}, 82 (2000), {\it ibidem} {\bf 573}, 652
(2000);  J.~R.~Ellis and S.~Lola, Phys.\ Lett.\ B {\bf 458}, 310 (1999)
P.~H.~Chankowski, A.~Ioannisian, S.~Pokorski and J.~W.~F.~Valle,
  Phys.\ Rev.\ Lett.\  {\bf 86}, 3488 (2001);
M.~C.~Chen and K.~T.~Mahanthappa,
Int.\ J.\ Mod.\ Phys.\ A {\bf 16}, 3923 (2001).
%%%%%mix enchancenc%%%%%%%%%%%%%%%%%%%%%%%%%%%%%%%
\bibitem{rad1} K. R. S. Balaji, A. Dighe, R. N. Mohapatra and M. K.
Parida, Phys. Rev. Lett. {\bf 84}, 5034 (2000); Phys. Lett. {\bf B
481}, 33 (2000); S.~Antusch and M.~Ratz,  JHEP {\bf 0211}, 010
(2002); R. N. Mohapatra, G. Rajasekaran and M. K. Parida, Phys.
Rev. {\bf D 69}, 053007 (2004);
\bibitem{renphases}
%%%%%%%%%%%%%%%%%%%%%%%%%%%%%%%%%%%%%%%%%%%%%%%%
  N.~Haba, Y.~Matsui and N.~Okamura,
  %``The effects of Majorana phases in three-generation neutrinos,''
  Eur.\ Phys.\ J.\ C {\bf 17}, 513 (2000).
%%%%%%%%%%rad%%  mass splitt%%%%%%%%%%%%%%%%%%%%%%%%%%%%%%%%
\bibitem{small} E.~J.~Chun,  Phys.\ Lett.\ B {\bf 505}, 155 (2001);
G. Bhattacharyya, A. Raichoudhuri and A. Sil,
Phys.Rev. {\bf D67}, 073004 (2003);
A.~S.~Joshipura, S.~D.~Rindani and N.~N.~Singh,
  %``Predictive framework with a pair of degenerate neutrinos at a high
%scale,''
  Nucl.\ Phys.\ B {\bf 660}, 362 (2003).



%%%%%%%%%%%%%%%%%%%%%%%%%%%%%%%%%%%%%%%%%%%%%%
\bibitem{Antusch:2002rr}
  S.~Antusch, J.~Kersten, M.~Lindner and M.~Ratz,
  %``Neutrino mass matrix running for non-degenerate see-saw scales,''
  Phys.\ Lett.\ B {\bf 538}, 87 (2002).
  %%CITATION = HEP-PH 0203233;%%




\end{thebibliography}
\end{document}